\documentclass[a4paper,11pt]{article}
\usepackage{amsmath}
\usepackage{a4wide}
\usepackage{multicol}
\usepackage{bm}
\usepackage{graphicx}
\usepackage{pstricks}
\usepackage{amsfonts}
\usepackage{mathrsfs}
\usepackage{cancel}
\usepackage{amssymb}
\usepackage{color}
\usepackage{upgreek}
\usepackage{caption}
\usepackage[numbers,super,sort&compress]{natbib}

\makeatletter
\DeclareRobustCommand\onlinecite{\@onlinecite}
\def\@onlinecite#1{\begingroup\let\@cite\NAT@citenum\citealp{#1}\endgroup}
\makeatother

\newcommand{\new}[1]{#1}

\newcommand{\pfrac}[2]{\frac{\partial{#1}}{\partial{#2}}}
\newcommand{\ppfrac}[2]{\frac{\partial^2{#1}}{\partial{#2}^2}}
\newcommand{\pppfrac}[2]{\frac{\partial^3{#1}}{\partial{#2}^3}}
\newcommand{\ppppfrac}[2]{\frac{\partial^4{#1}}{\partial{#2}^4}}

\newcommand{\refe}[1]{(\ref{#1})}
\newcommand{\mye}{\mbox{e}}

\newcommand{\tbf}[1]{\textbf{#1}}
\newcommand{\td}[1]{\tilde{#1}}
\newcommand{\infinity}{\infty}
\newcommand{\Ca}{\operatorname{Ca}}
\newcommand{\Rey}{\operatorname{Re}}
\newcommand{\BrG}{\bar{\rho}_G^s}
\newcommand{\BrS}{\bar{\rho}_S^s}
\newcommand{\Bax}{{x}}
\newcommand{\Baz}{{z}}
\newcommand{\Bat}{{t}}
\newcommand{\Bau}{{u}}
\newcommand{\Bav}{{v}}
\newcommand{\Bah}{{h}}
\newcommand{\Bap}{{p}}
\newcommand{\nabS}{\bm{\nabla}_{\hspace{-0.1cm}s}}
\newcommand{\pilfrac}[2]{\partial_{#2} {#1}}
\newcommand{\ppilfrac}[2]{\partial^2_{#2} {#1}}
\newcommand{\pppilfrac}[2]{\partial^3_{#2} {#1}}

\newcommand{\lrsq}[1]{\left[ {#1} \right]}

\newcommand{\lreval}[1]{\left. {#1} \right|}
\newcommand{\etal}{\textit{et~al.}}

\begin{document}


\title{Slip or not slip? A \new{methodical examination} of the interface formation model using two-dimensional droplet spreading on a horizontal planar substrate as a prototype system}

\author{David N. Sibley, Nikos Savva, and Serafim Kalliadasis
\\
\small{Department of Chemical Engineering, Imperial College London, London SW7 2AZ, UK}}
\date{\today}


\maketitle

\begin{abstract}
We consider the spreading of a thin two-dimensional droplet on a planar substrate as a prototype system to compare the contemporary model for contact line motion based on interface formation of Shikhmurzaev [Int. J. Multiphas. Flow {\bf 19}, 589 (1993)], to the more commonly used continuum fluid dynamical equations augmented with the Navier-slip condition. Considering quasistatic droplet evolution and using the method of matched asymptotics, we find that the evolution of the droplet radius using the interface formation model reduces to an equivalent expression for a slip model, where the prescribed \new{microscopic} dynamic contact angle has a velocity dependent correction to its static value. This result is found for both the original interface formation model formulation and for a more recent version, where mass transfer from bulk to surface layers is accounted for through the boundary conditions. Various features of the model, such as the pressure behaviour and rolling motion at the contact line, \new{and their relevance,} are also considered in the prototype system we adopt.
\end{abstract}

%
%

\section{Introduction}\label{sec:introduction}
A contact line is formed wherever two immiscible fluids, such as a liquid and a gas (often air) meet a solid surface. The wetting of a solid by a liquid occurs via a moving contact line in the gas/liquid/solid system, and is a crucial part of many natural and technological processes (see e.g. the review articles of de Gennes,\cite{deGennesrev} Blake,\cite{blake2006physics} and Bonn \textit{et al.}\cite{BonnEggers}). A model developed by Shikhmurzaev,\cite{Shikh93} \new{building on the earlier work of Bedeaux~\etal~\cite{BedeauxAlbanoMazur} on non-equilibrium thermodynamics for immiscible fluids,} is one of many put forward in the literature to understand this apparently simple situation. Since its inception the model has been used to describe a more general class of problems where interface formation occurs,\cite{ShikhBook} and as such, is termed the ``interface formation model".

The application of the usual hydrodynamic equations and fluid-mechanical modelling assumptions to the moving contact line leads to the nonexistence of a solution.\cite{ShikhSingualarities06} This issue was identified by Moffatt,\cite{Moffatt} and Huh and Scriven\cite{HuhScriv71} for a planar interface leading to the well-known problem of a nonintegrable shear-stress singularity. To alleviate the singularity, a variety of models and associated contributing phenomena have been proposed. These include allowing slip to occur at the solid surface, a rather popular model; accounting for the existence of a precursor film ahead of the contact line,\cite{BonnEggers} a popular model also; incorporating rheological effects such as shear-thinning;\cite{WeidnerSchwartz} incorporating intermolecular forces and considering the interface to be diffuse;\cite{PismenPomeau} or including evaporative fluxes\cite{ColinetRednikov} (for a review of some of these approaches in the context of the interface formation model see \S 3.4 of the monograph of Shikhmurzaev\cite{ShikhBook}). Savva and Kalliadasis\cite{SavvaPrecursorSlip} have recently demonstrated an equivalence between a selection of some of the most popular models, i.e. slip and precursor film models, in spreading situations \new{(building on the work of Pismen and Eggers\cite{PismenEggersSolv}),} so here we primarily assess the interface formation model in comparison to the popular slip model mentioned earlier, the Navier-slip condition~\cite{Navier} detailed in Sec.~\ref{sec:classical}.

\new{A feature, reported to be significant,\cite{ShikhLiqliqsolid97}} of the interface formation model over the usual hydrodynamic equations augmented with the Navier-slip boundary condition is that the \new{\textit{microscopic}} dynamic contact angle \new{(as opposed to the apparent contact angle at observable distances away from the contact line)} is determined as part of the solution. That no empirical relation between contact line velocity and contact angle is required, and additionally that the influence of the flow field on the contact angle is incorporated, makes the interface formation model worthy of analysis. \new{It should be noted, however, that this microscopic} \new{contact angle is determined through contact line conditions arising from a mechanical balance local to the contact line for the interface formation model, where similar reasoning leads to the imposition of the static Young's contact angle for the Navier-slip model at all times. Empirical relations may also be applied for the microscopic contact angle variation with slip models,\cite{HockingRival} but it is the fact that this is not required for the interface formation model that is of interest---especially as empirical relations will be based on experiments at observable distances from the contact line, but then enforced for the contact angle at the microscale. It is also worth mentioning here that precursor film models, where a very thin film covers the solid surface ahead of the macroscopic liquid/gas interface through the inclusion of a disjoining pressure (due to long-range intermolecular interactions), remove the contact line singularity as there is no longer a sharp contact line, but an apparent one. In these models this apparent contact angle is related to the model parameters, such as the Hamaker constant and film thickness, but no actual microscopic contact angle is formed. For further details of these and related aspects of the interface formation model, we refer the reader to Shikhmurzaev's earlier works\cite{Shikh93,ShikhLiqliqsolid97} and monograph.\cite{ShikhBook}}

However, as one might expect, the model has not remained free of controversy. There have been occasions where other authors have questioned its experimental validity (see Lindner-Silwester and Schneider,\cite{SilwesterOrig} noting the response of Shikhmurzaev\cite{ShikhResSilwester} and their reply\cite{SilwesterResShikh}), the assumptions underlying its formulation (see Eggers and Evans\cite{EggersOnShikh} and the response \cite{ShikhResEggers}), and the possibility of the requirement of an additional boundary condition at the contact line for consistency (suggested by Bedeaux\cite{BedeauxExtraCond} and Billingham\cite{Bill06}). \new{In a recent issue of the European Physical Journal - Special Topics, it is clear that the debate around the interface formation model remains very active, with strong opinions from a variety of authors from diverse scientific backgrounds published there.\cite{epj1,epj2,epj3,epj4,epj5,epj6,epj7,epj8,epj9,epj10,epj11,epj12}}

The two papers by Billingham,\cite{Bill06,BillGrav} \new{mentioned above,} for the steady motion of the contact line in a wedge and for a thin-film moving under gravity, give arguably the most careful analytical work with the interface formation model. We will consider the model in a similar framework but for a different problem, that of quasistatic droplet spreading. \new{This prototype system will also help to provide insight into the interface formation model, as we intend to scrutinise the model in light of the above opinions and debate.}

\new{Another feature of the model, much lauded by its advocates,} is that in a liquid/gas/solid system, the liquid is predicted to display rolling motion. This flow pattern has been experimentally observed by a variety of authors (see for instance Dussan V. and Davis\cite{DussanDavis} and Chen \textit{et al.},\cite{ChenCollSA,ChenRameGaroffJFM} \new{the latter with an inner limit at $\sim30\,\upmu$m from the contact line}), \new{but observations are on the micro- rather than the nano-scale, where significant differences between models will appear}. Rolling motion is built into the interface formation model through its formulation, whereas in the ``classical model'' the contact line instead slips over the solid. \new{Whilst this point will be evident from our analysis, and it is mentioned as a benefit of the interface formation model over others by Shikhmurzaev,\cite{ShikhBook} we reiterate that experimental observations of flow patterns at the microscale are in a region where slip models will also show similar flow patterns---albeit with the possibility of the stagnation point causing a slowing of the flow in the vicinity of the contact line for slip models.} We also highlight the experimental work of Savelski \textit{et al.},\cite{Cerro1} where it is suggested that rolling motion may instead occur in the gas phase with a splitting streamline present in the liquid (for particular values of the \new{apparent dynamic} contact angle and the viscosity ratio between the two fluids). These results are questioned by Shikhmurzaev\cite{ShikhLiqliqsolid97,ShikhBook} suggesting that the Reynolds number in such experiments must be based on the observable distance from the contact line and that the results do not invalidate the rolling pattern near the contact line where the Reynolds number tends to zero.

In earlier works of Shikhmurzaev (e.g.\cite{Shikh93,ShikhLiqliqsolid97}) and in the appraisals of Billingham,\cite{Bill06,BillGrav} mass exchange between the interfaces and the bulk was taken into account in the surface phase mass balance equations but neglected in the boundary conditions for the normal component of the bulk velocity. The neglected terms were found to be critical to the convergent flow of a Newtonian fluid near a free boundary, \cite{ShikhCusporCorner} and are included in the version given in Shikhmurzaev's monograph.\cite{ShikhBook} It is claimed that the earlier works considering dynamic wetting at small capillary numbers do not need to be revisited in light of the additional mass transfer contributions.\cite{ShikhCusporCorner,ShikhSingualarities06} Interestingly however, when considering the region close to the contact line we find in Sec.~\ref{sec:fullOadotchi0} that the low-velocity region predicted by slip models and preventing \new{nanoscale} rolling motion also occurs for the original formulation of the interface formation model. The additional terms in the modern formulation overcome this low-velocity region enabling rolling motion to take place, seen in Sec.~\ref{sec:chim1}. \new{{By considering the influence of the imposed interfacial boundary conditions, it may be concluded that a stagnation point at the contact line will occur for any model where mass transfer is not allowed. Diffuse interface methods, through the creation of an interfacial region of density variation, include mass transfer and should thus allow nanoscale rolling motion. A connection between diffuse interface methods and the interface formation model has been briefly suggested previously,\cite{PismenCusp,epj3} but a detailed comparison would be of interest}} \new{(although it lies beyond the scope of the analysis presented here).}

Whilst there have been a number of notable uses of the interface formation model,\cite{ShikhDrop,ShikhCapillaryBreakup,ShikhCRMecanique,ShikhCusporCorner,BlakeBrackeShikh,DecentCoalescenceDrops,DecentMicrodrop,Sprittles07ChemPat} there is still much room for further work. That there is not a much greater use of the model in the literature may \new{in part} be due to its comparative complexity, although \new{it is also likely that} the controversy discussed above \new{alongside questions over its physical basis have} dissuaded others to attempt detailed independent analyses of the model. Nevertheless, \new{whilst we have attempted to highlight the ongoing debate,} we believe that the model \new{does have a number of} advantages, and \new{that} further understanding of both the model and the questions raised about its formulation could aid in the long-standing issue of the moving contact line problem.

It is the intention of this work to consider the interface formation model in the simple yet illustrative situation of two-dimensional droplet spreading on a flat horizontal substrate utilising the long-wave approximation appropriate for thin droplets. The spreading of a droplet has been widely considered as a paradigmatic system to evaluate dynamic contact line motion by many authors. In particular, Hocking considered the competition of gravity and capillarity with Navier-slip,\cite{Hocking83} and Haley and Miksis extended this for other similar slip models.\cite{HaleyMiksis} The interface formation model has also been considered for droplet spreading\cite{ShikhDrop} with an extension for microdrops,\cite{DecentMicrodrop} but not within the long-wave approximation as is the case here. More specifically, by using this approximation for the problem of droplet spreading on a planar horizontal substrate, as in the framework of Hocking,\cite{Hocking83} we obtain a substantially reduced setting which is amenable to mathematical and numerical scrutiny -- thus allowing for direct comparisons between Navier-slip and interface formation models to be made with relative ease.

This work is outlined as follows. In the remainder of this section (in \ref{sec:classical}--\ref{sec:theifm}) we reprise the classical model and provide a summary of the interface formation model with some observations on the main points of its formulation. Then in Sec.~\ref{sec:appdropge} we detail the model equations and their forms in our droplet regime, followed by the nondimensionalization and long-wave assumption in Sec.~\ref{sec:nondim}. A matched asymptotic analysis is performed in Sec.~\ref{sec:secchim0} for the original interface formation model, with changes to the analysis for the modern formulation in Sec.~\ref{sec:chim1}. Our asymptotic matching procedure is verified through considering numerical solutions to the full problems, details in Sec.~\ref{sec:numerics}, with conclusions in Sec.~\ref{sec:conclusions}.

%
%

\subsection{Classical hydrodynamic model}
\label{sec:classical} For the classical model, the conditions on a flat solid boundary with unit normal and tangent vectors $\tbf{n}_S$, $\tbf{t}_S$ are, no-slip, $\tbf{u}\bm{\cdot}\tbf{t}_S=0$, and no-penetration, $\tbf{u}\bm{\cdot}\tbf{n}_S=0$, where $\tbf{u}$ is the bulk fluid velocity. As detailed earlier, enforcing the no-slip condition leads to the nonexistence of a solution to the moving contact line problem. To alleviate this difficulty the classical model invokes the popular Navier-slip condition
\begin{equation}
 \tbf{u}\bm{\cdot}\tbf{t}_S = \frac{\beta_{NS}}{\mu} \tbf{n}_S \bm{\cdot} \bm{\sigma} \bm{\cdot}\tbf{t}_S,
\end{equation}
where $\mu$ is the fluid viscosity, $\beta_{NS}$ is the coefficient of slip, and $\bm{\sigma}=-P\tbf{I}+\tbf{T}$ is the total stress tensor, where $P$ is the pressure, $\tbf{I}$ is the identity tensor, and $\tbf{T}=2\mu\tbf{D}=\mu\left[(\bm{\nabla}\tbf{u})+(\bm{\nabla}\tbf{u})^\mathrm{T}\right]$ is the extra-stress tensor. On a free surface between a liquid and a gas with unit normal and tangent vectors $\tbf{n}_G$, $\tbf{t}_G$, we have
\begin{equation}
 \frac{Df}{DT} = \pfrac{f}{T}+\tbf{u}\bm{\cdot\nabla}f = 0 , \qquad
 \tbf{n}_G\bm{\cdot}\tbf{T}\bm{\cdot}\tbf{t}_G = -\nabS\sigma_{NS}\bm{\cdot}\tbf{t}_G, \qquad
 P_G-P+\tbf{n}_G\bm{\cdot}\tbf{T}\bm{\cdot}\tbf{n}_G = \sigma_{NS}\nabS\bm{\cdot} \tbf{n}_G,
\end{equation}
which are the kinematic condition, the tangential stress equation, and the normal stress equation, where $\sigma_{NS}$ is the surface tension on the liquid-gas interface $f(r,\theta)=0$, $P_G$ is the pressure of the gas, and $T$ is time. The tangential stress equation gives rise to the Marangoni effect when a surface tension gradient is present, caused by gradients in temperature or variation in chemical composition at the interface. $\nabS \sigma_{NS}$ is the surface gradient of $\sigma_{NS}$. For scalar field $f$ and vector field $\tbf{f}$, we define
\begin{equation}\label{eq:sgrad}
 \nabS f = (\tbf{I}-\tbf{n}\otimes\tbf{n})\bm{\cdot}\bm{\nabla}f,\qquad
 \nabS\bm{\cdot}\tbf{f} = \mbox{tr}\left( (\tbf{I}-\tbf{n}\otimes\tbf{n})\bm{\cdot}\bm{\nabla} \tbf{f} \right),
\end{equation}
where on the right hand side, $\bm{\nabla}$ is the usual gradient operator, and $\tbf{n}\otimes\tbf{n}$ is the dyadic product of the normal vector $\tbf{n}$ with itself.\cite{slatteryitp}

Finally, at the contact line there is no way of determining the \new{microscopic} dynamic contact angle $\theta_d$ from the classical model. As such it must be prescribed \new{to provide a boundary condition at the contact line}, either by assuming that it is always equal to the static contact angle, $\theta_d=\theta_s$, or by imposing a relationship on the \new{microscopic} dynamic contact angle in terms of the contact line velocity. The interface formation model by comparison has the \new{microscopic} dynamic contact angle determined as part of the solution \new{via the Young equation}, as mentioned earlier, so that no empirical relation is required.

%
%

\subsection{The interface formation model}
\label{sec:theifm}

\begin{figure}[ht]
\centering
\includegraphics{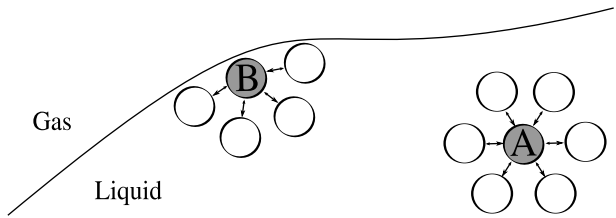}
\caption{The density variation close to an interface, and the resulting surface tension. Fluid particle A away from the interface is surrounded by other fluid particles, whereas the fluid particle B near the interface feels the presence of fewer fluid particles and is thus in an unfavourable state. It is this which gives rise to the surface tension on continuum mechanical scales.}
\label{fig:surfacetension}
\end{figure}

\begin{figure}[ht]
 \centering
    \includegraphics{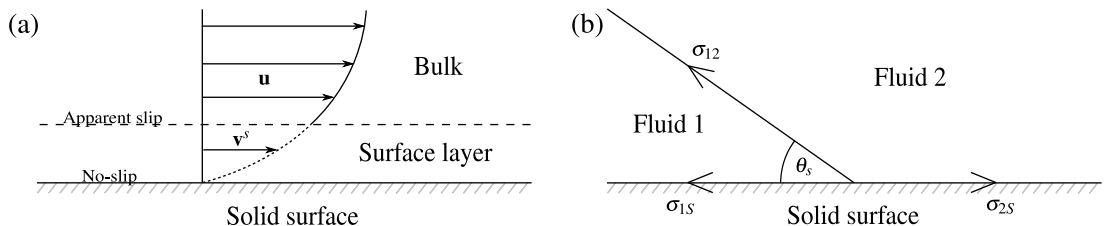}
    \caption{(a) Sketch of the surface layer with velocity at the solid surface satisfying no-slip. This surface layer is taken with zero thickness in the continuum approximation so that the bulk velocity $\tbf{u}$ undergoes apparent slip at the boundary. (b) A Young's force diagram showing the forces acting at a static contact line between two fluids and a solid substrate, with the three phases meeting at static contact angle $\theta_S$ (see Billingham\cite{Bill06}).}
\label{fig:noslipslip}
\end{figure}

There are two interrelated mechanisms key to the formulation of the interface formation model:
\begin{enumerate}
\item Surface layers: In a region near to an interface between a liquid and another substance, liquid molecules are not fully surrounded by other liquid molecules as they would be in the bulk due to the proximity of the other material. As such they lose a number of cohesive interactions causing an uneven force on the liquid molecules and giving rise to a thin surface layer where the liquid density differs from the bulk (Billingham\cite{Bill06} quotes a typical thickness of $10^{-10}\mbox{m}$). This variation in density viewed on a macroscopic scale, appears as the surface tension of the liquid, see Fig.~\ref{fig:surfacetension}. This can also be thought of as liquid molecules near the surface being in an unfavourable energy state and the surface tension being a direct measure of this energy shortfall per unit area.\cite{deGennesbook}

Consider a surface layer with surface velocity $\tbf{v}^s$. The conditions on either side of the layer may differ---for instance the surface layer between liquid and solid may exhibit no-slip on the solid-facing side (in accordance with the classical model), but on the liquid-facing side may slip, see Fig.~\ref{fig:noslipslip}(a). The continuum mechanical limit of these surface layers being of zero thickness is then taken, and thus the surface properties influence the bulk simply by modifying the boundary conditions applied at each interface. We summarise the implication of surface layers as follows:
    \begin{itemize}
        \item Density in surface layer differs from that of the bulk due to presence of other material,
        \item Density variation gives rise to surface tension,
        \item Allows different conditions to be applied on each side of the surface layer---no-slip is then satisfied for the fluid particles actually next to the solid, with a slip condition appearing as the boundary condition due to averaging throughout the surface layer.
    \end{itemize}
The surface layers have the properties of surface velocity $\tbf{v}^s$, surface tension $\sigma^s$, surface density $\rho^s$, equilibrium surface density $\rho^s_e$ and a surface density associated with zero surface tension $\rho^s_{(0)}$. These will be differentiated between the solid-liquid and the gas-liquid interfaces with subscripts $S$ and $G$ respectively.

\item Surface tension relaxation: Consider a static contact line as in Fig.~\ref{fig:noslipslip}(b), where the surface tensions associated with the interfaces between fluids 1 and 2 and the solid surface are $\sigma_{12}$, $\sigma_{1S}$ and $\sigma_{2S}$. In this situation, the horizontal force balance is given by the Young equation:
\begin{equation}
 \sigma_{12}\cos\theta_s + \sigma_{1S} = \sigma_{2S}. \label{eq:lwIFM_young}
\end{equation}
[It is noteworthy, that beside the Young equation, which reflects the mechanical equilibrium in a direction parallel to the wall at the contact line, there exists a relation in the normal direction\cite{Henderson,PereiraKalliadasis} but its implications are beyond the scope of this study.] For a moving contact line \refe{eq:lwIFM_young} also holds as any momentum fluxes due to the motion of the interfaces are negligible (see \S 4.3.5.2 of Ref.~\onlinecite{ShikhBook}), but in this case with dynamic values of the surfaces tensions. When in motion, the \new{microscopic} dynamic contact angle varies from its static value causing either $\sigma_{12}$, $\sigma_{1S}$, $\sigma_{2S}$ or a combination to also deviate from their static values. Discussion about the Young equation including some recent disputes is also given in \S 2.4.4 of the monograph of Shikhmurzaev.\cite{ShikhBook}

Material flux through the contact line is caused by the rolling motion of the fluid. As the properties of the liquid-gas and liquid-solid interfaces will not be the same in general, then as a fluid particle is transferred from the gas interface to become part of the solid interface, some reorganisation of the molecules will be required. In particular, a fluid particle is associated with the surface tension $\sigma_{12}$ when on the gas interface, but after moving through the contact line must then be associated with the surface tension $\sigma_{1S}$. The crucial assumption here is that this relaxation of the surface tension to the equilibrium existing far from the contact line happens in a finite time, $\tau$, rather than instantaneously, \cite{blake2006physics,Bill06} \new{this assumption being vital for the interface formation model as an instantaneous relaxation will reduce the model to that of classical Navier-slip.} 

\end{enumerate}

It is important to point out that in the following we have the unit normal at both solid and gas interfaces $\tbf{n}_S$ and $\tbf{n}_G$ pointing into the liquid as defined by Shikhmurzaev,\cite{ShikhBook} where we also refer the reader for further information about the interface formation model and for details of its derivation.

\paragraph*{The interface formation model equations}

We now consider the equations of the interface formation model. 
First consider the flow in the bulk liquid, which is governed by the usual Navier--Stokes equations for a viscous incompressible Newtonian fluid
\begin{equation}
 \bm\nabla \bm{\cdot}\tbf{u} = 0, \quad \rho\left(\pfrac{\tbf{u}}{T}+\tbf{u}\bm{\cdot}\bm{\nabla}\tbf{u}  \right)= - \bm\nabla P+\mu\bm\nabla^2\tbf{u},
\end{equation}
where $\rho$ is the bulk density, other variables are as for the classical model and gravitational effects are assumed to be negligible. We now consider the solid-liquid interface:
\begin{gather}
\tbf{v}^s_S\bm{\cdot}\tbf{n}_S=0, \label{eq:sbc1}\\
\label{eq:sbc2}\refstepcounter{equation}\tag{\theequation \ a,b}
\tbf{t}_{xy}+\frac{1}{2}\nabS\sigma_S^s=\beta\tbf{u}_{||}, \qquad
\rho(\tbf{u}-\tbf{v}_S^s)\bm{\cdot}\tbf{n}_S=\chi_{m}\frac{\rho_S^s-\rho_{Se}^s}{\tau},\\
\label{eq:sbc3}\refstepcounter{equation}\tag{\theequation \ a,b,c}
\pfrac{\rho_S^s}{T}+\nabS\bm{\cdot}(\rho_S^s\tbf{v}_S^s) = -\frac{\rho_S^s-\rho_{Se}^s}{\tau},
\qquad
\tbf{v}_{S||}^s = \frac{1}{2}\tbf{u}_{||}+\alpha\nabS\sigma_S^s,
\qquad
\sigma_S^s=\gamma(\rho_{(0)S}^s-\rho_S^s),
\end{gather}
where $\tbf{u}_{||}=(\tbf{I}-\tbf{n}_S\otimes\tbf{n}_S)\bm{\cdot}\tbf{u}$ and $\tbf{v}_{S||}^s=(\tbf{I}-\tbf{n}_S\otimes\tbf{n}_S)\bm{\cdot}\tbf{v}_{S}^s$ are the tangential projections of the bulk and surface velocities at the solid surface, and $\tbf{t}_{xy}=2\mu\tbf{n}_S\bm{\cdot}\tbf{D}\bm{\cdot}(\tbf{I}-\tbf{n}_S\otimes\tbf{n}_S)$
is the shear stress exerted by the bulk fluid on the interface. Selecting $\chi_m=0$ or $\chi_m=1$ determines whether mass transfer between the surface layer and bulk is included throughout the surface layer equations, giving either the original or modern formulation of the interface formation model. These equations are for a static solid substrate so there is no solid velocity contribution, in comparison to the equations in the monograph of Shikhmurzaev.\cite{ShikhBook} We choose to work in a frame of reference of a static observer (such as also taken by Billingham\cite{BillGrav} and Shikhmurzaev\cite{ShikhDrop}).

The equations above have been grouped to broadly indicate where they apply, if imagining the surface with a thickness for illustrative purposes. Equation \refe{eq:sbc1} applies at the solid-facing side of the surface layer, the condition describing the usual no-penetration of the liquid. Equations (7) apply at the liquid-facing side, being a generalised Navier-slip condition and a condition on the normal component of the bulk velocity (with mass transfer between bulk and surface layer included or neglected through the choice of $\chi_m$). Equations (8) apply throughout the surface layer, giving the mass balance, a surface velocity relation (being a jump in tangential velocities due to the surface tension gradient), and a relationship between surface tension and surface density---a linear relationship is assumed, although Shikhmurzaev gives suggestions as to how this may be generalised if required.\cite{ShikhBook} If there was no surface tension gradient, the surface velocity relation would suggest the surface velocity is the average of the bulk and solid velocities, whereas with surface tension gradient, the effect is similar to Darcy's law for the pressure drop in a porous pipe.\cite{batchelor}

We note that $\alpha$, $\beta$ and $\gamma$ are parameters of the surface (typical values quoted in Billingham\cite{Bill06}, \new{see also Table~1}), with $\alpha$ corresponding to the response of the interface to surface tension gradients, $\beta$ to the coefficient of sliding friction and $\gamma$ to the inverse surface layer compressibility. The partial derivative of a surface variable, such as in (8a), needs to be considered with care, but any differences between definitions are negligible in the long-wave approximation.\cite{PereiraTransportEq}

Next we consider the gas-liquid interface:
\begin{gather}
 \frac{Df}{DT}=\pfrac{f}{T}+\tbf{v}_G^s\bm{\cdot}\bm{\nabla}f=0,\label{eq:lbc1}
\\
\label{eq:lbc2}\refstepcounter{equation}\tag{\theequation \ a,b,c}
P_G-P+\tbf{n}_G\bm{\cdot}\tbf{T}\bm{\cdot}\tbf{n}_G=\sigma_G^s\nabS\bm{\cdot} \tbf{n}_G, \qquad
\tbf{t}_{xy}=-\nabS\sigma_G^s, \qquad
\rho(\tbf{u}-\tbf{v}_G^s)\bm{\cdot}\tbf{n}_G=\chi_m\frac{\rho_G^s-\rho_{Ge}^s}{\tau},
\\
\label{eq:lbc3}\refstepcounter{equation}\tag{\theequation \ a,b,c}
\pfrac{\rho_G^s}{T}+\nabS\bm{\cdot}(\rho_G^s\tbf{v}_G^s) = -\frac{\rho_G^s-\rho_{Ge}^s}{\tau}, \qquad
\frac{4\beta}{1+4\alpha\beta} (\tbf{v}_{G||}^s-\tbf{u}_{||})=\nabS\sigma_G^s, \qquad
\sigma_G^s=\gamma(\rho_{(0)G}^s-\rho_G^s),
\end{gather}
where similarly $\tbf{u}_{||}=(\tbf{I}-\tbf{n}_G\otimes\tbf{n}_G)\bm{\cdot}\tbf{u}$ and $\tbf{v}_{G||}^s=(\tbf{I}-\tbf{n}_G\otimes\tbf{n}_G)\bm{\cdot}\tbf{v}_{G}^s$ are the tangential projections of the bulk and surface velocities, and $\tbf{t}_{xy}=2\mu\tbf{n}_G\bm{\cdot}\tbf{D}\bm{\cdot}(\tbf{I}-\tbf{n}_G\otimes\tbf{n}_G)$ is the shear stress. Equation \refe{eq:lbc1} gives the usual kinematic condition, applying on the gas-facing side of the surface, the free surface being associated with the surface velocity $\tbf{v}_G^s$. Equations (10) apply at the liquid-facing side and are the normal and tangential stress conditions \new{(where viscous gas stresses are assumed negligible, as in many other works with both interface formation model\cite{BillGrav} and classical models\cite{deGennesrev,BonnEggers,Hocking83})}, and a condition on the normal component of the bulk velocity. Unlike the classical formulation the gradient of surface tension plays a role here purely based on the flow of the fluid---with no variation in temperature or chemical composition required (Marangoni effect). Equations~(11) finally are the analogues of (8) above, \new{with the unusual factor in (11b) readily derived from general (rather than specific to solid-liquid or liquid-gas) surface equations in the form of (7a), (8b) and (10b), see Ref.~\onlinecite{ShikhBook} for details of the derivation}. It should be noted here that $\alpha$, $\beta$, $\gamma$ and $\tau$ have been used for both surface layers and for simplicity are assumed to be equal for both (consistent with other works with this model \cite{Bill06,BillGrav,ShikhDrop}).

Finally, we require conditions at the contact line. It is postulated in Refs.~\onlinecite{Shikh93,ShikhBook} that there are two conditions at the contact line.
\begin{itemize}
 \item Conservation of mass through the contact line:
 \begin{equation}
  \rho_G^s(\tbf{v}_G^s-\tbf{U}_{CL})\bm{\cdot}\tbf{t}_G +  \rho_S^s(\tbf{v}_S^s-\tbf{U}_{CL})\bm{\cdot}\tbf{t}_S + Q = 0,\label{eq:comCL}
 \end{equation}
where we have included the contribution from the contact line velocity, $\tbf{U}_{CL}$ since (as mentioned earlier) we are in the frame of reference of a static observer. Equation \refe{eq:comCL} states that the mass flowing into the contact line from the gas-liquid surface equals the amount flowing out into the solid-liquid surface, with the possibility of mass transfer to the bulk with non-zero $Q$. It is assumed in most previous works that $Q=0$, and we make the same assumption here.
\item A force balance at the contact line, from \refe{eq:lwIFM_young}:
\begin{equation}
\label{eq:cmomCL}
\sigma_{G}^s\cos\theta_d+\sigma_{S}^s = \sigma_{Ge}^s\cos\theta_s+\sigma_{Se}^s,
\end{equation}
where the assumption is made that the surface tension associated with the solid-gas interface, $\sigma^{s}_G$, does not vary from its static value as the contact line moves.\cite{BillGrav,DecentMicrodrop} \\ Shikhmurzaev comments that $\sigma^s_G$ is negligible for all contact angles except if close to $180^\circ$, leading to the same condition above.\cite{ShikhBook} As such there appears to be unanimous agreement on this boundary condition as we use it.
\end{itemize}
A third condition is found necessary by Billingham\cite{Bill06} and derived by Bedeaux\cite{BedeauxExtraCond} through consideration of the entropy production rate at the contact line\new{, which relies on the ability to treat the contact line as a separate thermodynamic system---with associated properties such as line tension. Shikhmurzaev, in \S 4.3.5.3 of his monograph\cite{ShikhBook} puts forward his point of view regarding this treatment, suggesting that a line phase cannot exist whilst remaining in the continuum approximation, referencing other authors such as de Gennes.\cite{deGennesbook} There remains, however, much debate and disagreement over this point and if one follows the \new{recent} work of Bedeaux\new{\cite{BedeauxExtraCond}}, then the treatment of the line phase considering conservation of energy arguments leads to the condition}
\begin{equation}
\label{eq:coeCL}
 \rho_S^s(\tbf{v}_S^s-\tbf{U}_{CL})\bm{\cdot}\tbf{t}_S = U_0 (\rho_S^s - \rho_{Se}^s) + U_{G0}(\rho_G^s - \rho_{Ge}^s),
\end{equation}
where $U_0$ and $U_{G0}$ are constants with dimensions of velocity and assumed to be equal for simplicity (as assumed by Billingham,\cite{BillGrav} \new{with \refe{eq:coeCL} given as found there}). This condition states that the flux through the contact line is driven by the deviation of the surface layer densities from equilibrium.

\new{As mentioned earlier, a full derivation of the interface formation model equations can be found in Ref.~\onlinecite{ShikhBook}. We highlight that equations \refe{eq:sbc1}, \refe{eq:lbc1} and \refe{eq:comCL} arise through conservation of mass, and the normal and tangential stress balances in (10 a,b) as well as contact line condition \refe{eq:cmomCL} arise from conservation of momentum. The remaining surface equations are all derived from requiring nonnegative entropy production in the surface phases, with results from mass and momentum conservation also applied in certain circumstances. Whilst this connection to entropy production is not immediately apparent from the forms of the equations, the derivation connects terms involving the entropy to the surface chemical potential $\mu^s$, arising from the Gibbs relation. The interface formation model equations (7b) and (10c), and consequently (8a) and (11a) are then obtained through the assumption that the deviations of the surface density from equilibrium are small so that the chemical potential may be expanded in a Taylor series about $\rho_{e}^s$, the equilibrium surface density (see \S 4.3.2 of Ref.~\onlinecite{ShikhBook}, or equation (21) of Ref.~\onlinecite{Shikh93}), and thus
\begin{equation}
 \mu^s = \mu^s_e + \lreval{\frac{d \mu^s}{d \rho^s}}_{\rho^s=\rho^s_e}(\rho^s-\rho^s_e),
\end{equation}
so that the governing equations are now written in terms of the surface density --- with
$d_{\rho^s} \mu^s(\rho^s_e)$ being absorbed into a surface coefficient becoming the relaxation timescale $\tau$.
It is this treatment that then motivates a similar consideration of the line phase of Bedeaux by Billingham,\cite{BedeauxExtraCond} and why although condition \refe{eq:coeCL} may initially appear more like a momentum balance, it does in fact arise from energy considerations, effectively} \new{suggesting that the flow through the contact line is driven by a difference in the chemical potentials of the interfaces there. We reiterate that this additional condition requires treating the contact line region as a separate thermodynamic entity, which Shikhmurzaev disagrees with, suggesting that there are no singularly strong forces when going down a dimension from surface to line in an analogous way to the asymmetric intermolecular forces when going from bulk to surface, shown in Fig.~\ref{fig:surfacetension}, which give rise to the surface tension.\cite{ShikhBook}}

\new{Before proceeding with the investigation of the interface formation model equations for the droplet spreading problem of interest, we will make some further remarks on the model. There are a number of ``non-classical'' ingredients which are included within the model, having arisen through its derivation using non-equilibrium thermodynamics. In particular, these include generalised slip (in (7a)), mass transfer between bulk and surface layers (in (7b), (8a), and (10c), (11a)), and flow-induced Marangoni effect (within (7a), (8b), and (10b), (11b)). The model, as mentioned previously, will reduce to the Navier-slip model if the relaxation of the surface tension is instantaneous. The slip terms remove the moving contact line singularity, although the subtle distinction is that (as highlighted in Fig. \ref{fig:noslipslip}) the slip occurs between bulk and surface layers rather than at the solid surface. The mass transfer and Marangoni effect terms are then intrinsically linked as it is the finite time surface tension relaxation which induces the Marangoni effect, and the surface tension varies dynamically due to the variable mass in the surface layers. These features are critical to the ability of the interface formation model to allow and determine the dynamic variation in the microscopic contact angle, as it evolves with the dynamic surface tensions through the Young equation \refe{eq:lwIFM_young}.}

\new{It is certainly of interest to consider whether one surface may be described classically, with the other using elements of the interface formation model. The assumption made here that the model parameters $\alpha$, $\beta$, $\gamma$ and $\tau$ are equal for both surface layers would need to be relaxed, and further experimental work to determine the relaxation time $\tau$ for different surfaces is required to justify the assumption that it is negligibly small. A previous study by Monnier and Witomski\cite{MonnierWitomski} considered the interface formation model numerically for a plunging tape. They assumed that the liquid-gas surface tension was always at its equilibrium value, so some analysis of a simplified model has been undertaken. The authors there only make this assumption for simplicity, however, and state that their results are only a first step (suggesting that the model is too simplified to interpret results from a mechanical point of view), with development of the full model in progress.}

%
%

\section{Application to droplet spreading}
\label{sec:appdrop}
\subsection{Governing equations}
\label{sec:appdropge}
\begin{figure}[ht]
    \centering
    \includegraphics{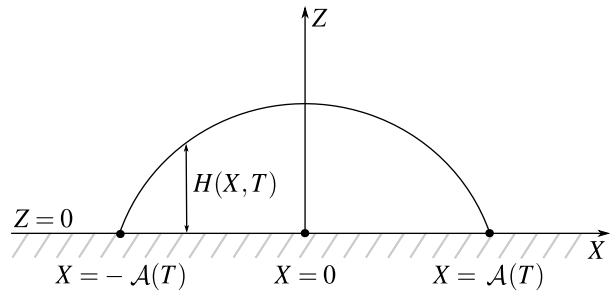}
    \caption{Symmetric two-dimensional droplet of thickness $Z=H(X,T)$ with contact lines at $X=\pm\mathcal{A}(T)$ and coordinate axes chosen to be parallel and perpendicular to the planar solid substrate.}
\label{fig:drop}
\end{figure}

We now formulate the governing equations for our prototype system, a two-dimensional droplet on a horizontal planar substrate. $(X,Z)$-axes are chosen parallel and perpendicular to the planar solid substrate, respectively, with $X=0$ the centre line of the droplet, $X=\pm \mathcal{A}(T)$ the location of the contact lines, $Z=H(X,T)$ the height of the droplet, and $\tbf{u}=(U,V)$, see Fig.~\ref{fig:drop}.
The bulk motion of the incompressible Newtonian fluid is given by the continuity equation, $\pilfrac{U}{X} + \pilfrac{V}{Z} = 0$, and conservation of momentum equations,
\begin{subequations}
\begin{align}
\rho\left(\pfrac{U}{T}+U\pfrac{U}{X}+V\pfrac{U}{Z}  \right) &= -\pfrac{P}{X}+\mu\left(\ppfrac{U}{X}+\ppfrac{U}{Z}\right),
\\
\rho\left(\pfrac{V}{T}+U\pfrac{V}{X}+V\pfrac{V}{Z}  \right) &= -\pfrac{P}{Z}+\mu\left(\ppfrac{V}{X}+\ppfrac{V}{Z}\right).
\end{align}
\end{subequations}
The solid-liquid interface has unit normal $\tbf{n}_S=(0,1)$ into the liquid, $\tbf{v}^s_S=(u^s_S,0)$ from \refe{eq:sbc1}, and
\begin{gather}
\mu\left( \pfrac{U}{Z} + \pfrac{V}{X} \right)+\frac{1}{2}\pfrac{\sigma_S^s}{X}=\beta U,\qquad
\rho V = \chi_m\frac{\rho_S^s-\rho_{Se}^s}{\tau},\nonumber\\
\pfrac{\rho_S^s}{T}+\pfrac{(\rho_S^s {u}_S^s)}{X} = -\frac{\rho_S^s-\rho_{Se}^s}{\tau},\qquad
{u}_S^s = \frac{1}{2}{U}+\alpha\pfrac{\sigma_S^s}{X},\qquad
\sigma_S^s=\gamma(\rho_{(0)S}^s-\rho_S^s).\label{eq:2dsbc6}
\end{gather}
The gas-liquid interface has normal pointing into the liquid droplet
\begin{equation}
\tbf{n}_G=\frac{1}{\sqrt{\mathcal{H}}}\left(\pilfrac{H}{X},-1\right), \quad \mbox{giving} \quad
\nabS\bm{\cdot}\tbf{n}_G=\frac{\ppilfrac{H}{X}}{\mathcal{H}^{3/2}}
,
\end{equation}
where $\mathcal{H}=1+(\pilfrac{H}{X})^2$ is written for simplicity, and then
\begin{equation}
 \tbf{n}_G\bm{\cdot}\tbf{T}\bm{\cdot}\tbf{n}_G =
\frac{2\mu}{\mathcal{H}}\left[ \left(\mathcal{H}-2\right)\pfrac{U}{X} - \left( \pfrac{U}{Z}+\pfrac{V}{X} \right)\pfrac{H}{X} \right]
,\label{eq:normst}
\end{equation}
so that the gas surface equations become
{\allowdisplaybreaks
\begin{subequations}
\begin{gather}
 \pfrac{H}{T}+u_G^s\pfrac{H}{X}=v_G^s,
\\
P_G-P+\tbf{n}_G\bm{\cdot}\tbf{T}\bm{\cdot}\tbf{n}_G=\sigma_G^s\nabS\bm{\cdot} \tbf{n}_G,\\
\left(\mathcal{H}-2\right)\left( \pfrac{U}{Z}+\pfrac{V}{X} \right)+4\pfrac{U}{X}\pfrac{H}{X}  =-\frac{\sqrt\mathcal{H}}{\mu}\pfrac{\sigma_G^s}{X},
\\
(U-{u}_G^s)\pfrac{H}{X}-V+v_G^s=\chi_m\sqrt\mathcal{H}\frac{\rho_G^s-\rho_{Ge}^s}{\rho\tau},\label{eq:2dlbc4}\\
\pfrac{\rho_G^s}{T}+ \frac{1}{\mathcal{H}}\left[
{ \pfrac{}{X}(\rho_G^s{u}_G^s) + \pfrac{H}{X}\pfrac{}{X}(\rho_G^s{v}_G^s)} \right]
= \frac{\rho_{Ge}^s-\rho_G^s}{\tau},\\\label{eq:2dlbc6}
\frac{4\beta}{1+4\alpha\beta}
\left[{u}_{G}^s-U+(v_G^s-V)\pfrac{H}{X}\right]
=\pfrac{\sigma_G^s}{X},\\
\sigma_G^s=\gamma(\rho_{(0)G}^s-\rho_G^s),\label{eq:2dlbc7}
\end{gather}
\end{subequations}
}%
where $\tbf{v}^s_G=(u^s_G,v^s_G)$. The surface gradient and surface divergence from \refe{eq:sgrad} have been used, and we confirm $\tbf{n}_G\bm{\cdot}\nabS\sigma = 0$, $\tbf{n}_G\bm{\cdot}\nabS\tbf{n}_G=\bm{0}$, and $\tbf{n}_G\bm{\cdot}\nabS\tbf{v}_G^s=\bm{0}$, as would be expected. Finally at the contact line
\begin{gather}
  \rho_G^s\left[({u}_G^s-{U}_{CL})\cos\theta_d + {v}_G^s\sin\theta_d \right] +  \rho_S^s({u}_S^s-{U}_{CL}) = 0,
\nonumber\\
\sigma_{G}^s\cos\theta_d+\sigma_{S}^s = \sigma_{Ge}^s\cos\theta_s+\sigma_{Se}^s,
\nonumber\\
 \rho_S^s({u}_S^s-{U}_{CL})  = U_0 (\rho_S^s - \rho_{Se}^s + \rho_G^s - \rho_{Ge}^s),\label{prenondim:clcond}
\end{gather}
as $\tbf{U}_{CL}=(U_{CL},0)$. Noting that the free surface moves at velocity $-\tbf{v}^s_G\bm{\cdot}\tbf{n}_G$, then we also have:
\begin{equation}
 U_{CL} = -\frac{\tbf{v}^s_G\bm{\cdot}\tbf{n}_G}{\sin\theta_d} = \frac{v_G^s-u_G^s\pilfrac{H}{X}}{\sin\theta_d\sqrt\mathcal{H}}.
\end{equation}
It is possible to simplify the above equations by removing excess variables, details are recorded in the Appendix. The next step will be to nondimensionalise these equations and determine which terms can be neglected under the long-wave approximation. We note that for our specific droplet spreading problem we have conservation of mass suggesting
\begin{equation}
\pfrac{}{T}\int_0^{\mathcal{A}(T)}\int_0^{H(X,T)}\rho dZ dX = \frac{\chi_m}{\tau}\int_0^{\mathcal{A}(T)}
(\rho_S^s - \rho_{Se}^s + \rho_G^s - \rho_{Ge}^s ) dX, \label{eq:cofmdim}
\end{equation}
taking into account the mass flux into the surface layers. There will also be a number of additional constraints imposed to preserve symmetry, relate \new{microscopic} contact angle to the slope of the free surface etc., but these will be described later.

%
%

\subsection{Nondimensionalization}
\label{sec:nondim}
We proceed by defining nondimensional variables using the following substitutions:
\begin{gather}
 X = L\Bax, \quad Z = \epsilon L\Baz, \quad T = \frac{L}{\mathcal{U}}\Bat, \quad H = \epsilon L\Bah,\quad
 U = \mathcal{U}\Bau, \quad
V = \epsilon \mathcal{U}\Bav, \quad U_{CL} = \mathcal{U}\bar{U}_{CL}, \nonumber\\
 P = \frac{\mu \mathcal{U}}{\epsilon^2 L} \Bap, \quad \rho_S^s = \rho_{Se}^s +\frac{\mu \mathcal{U}}{\epsilon \gamma} \bar{\rho}_S^s, \quad
 \rho_G^s = \rho_{Ge}^s +\frac{\mu \mathcal{U}}{\epsilon \gamma} \bar{\rho}_G^s,
\end{gather}
where lower-case letters and barred variables are nondimensional quantities. $L$ is a typical length scale in the $x$-direction such that $x=\pm\mathcal{A}/L=\pm a$ is the location of the contact lines, $\mathcal{U}$ is a typical horizontal velocity scale, and $\epsilon$ is a small parameter allowing for the long-wave approximation (often referred to as the ``film parameter" in thin-film studies, e.g. Ref.~\onlinecite{Kalliadasis_etal}). Here we assume $\epsilon=\theta_s$, with $\theta_d=O(\epsilon)$. Our governing equations contain the following dimensionless numbers
\begin{gather}
 \Rey = \frac{\rho \mathcal{U} L}{\mu}  \approx \frac{7\times10^7 L \epsilon^3}{3} , \quad
 \bar{\beta} = \frac{\beta L \epsilon}{3\mu} \approx \frac{10^{10} L \epsilon}{3} , \quad
 \bar{R}_0 = \frac{3\mathcal{U}L}{\gamma\tau\Rey\epsilon^2}  \approx \frac{3\times10^{-4}}{2\epsilon^2} ,
\nonumber\\
 \bar\alpha=\alpha\beta\approx \frac{1}{12},
\quad
 \bar{\tau} = \frac{3\tau \mathcal{U}\lambda}{L\epsilon^3} \approx \frac{2\times 10^{-6}\epsilon}{3L}, \nonumber\\
 \lambda = \frac{\epsilon^4 \gamma \rho_{Ge}^s}{9\mu\mathcal{U}} \approx \frac{20}{21}\epsilon,\quad
 \breve{\rho} = \frac{\rho_{Se}^s}{\rho_{Ge}^s}\approx 1, \quad
 \Ca_0 = \frac{\mu U_0}{\sigma_{Ge}^s} \approx 1.4 \times 10^{-6},
\label{eq:paramsizes}
\end{gather}
where certain scalings have been chosen to achieve the fullest balance in the equation for the pressure jump at the gas surface, and $\mathcal{U}=\sigma_{Ge}^s\epsilon^3/(3\mu)=O(70\epsilon^3/3)$. The nondimensionalization is motivated by that of Billingham\cite{Bill06} with additional scalings based on Hocking\cite{Hocking83} to allow for direct comparison to the Navier-slip case, \new{and hence why certain numerical values in \refe{eq:paramsizes} are present}. Approximate parameter values are based on those given in Table 1 of Billingham\cite{Bill06} for water at room temperature, \new{and relevant quantities are also included in Table \ref{table:physvals} of the present communication for self-consistency}.

\begin{minipage}{0.9\linewidth}
\centering
\captionof{table}{Physical parameters estimated for water at room temperature, values from Billingham, Ref.~\onlinecite{Bill06}.}\label{table:physvals}
\begin{tabular}{ccc}
\tbf{Physical quantity} & \tbf{Symbol} & \tbf{Estimated typical value} \\\hline\hline
inverse surface layer compressibility & $\gamma$ & $2\times10^6 \ \mbox{m}^2 \mbox{s}^{-2}$\\
surface layer relaxation timescale & $\tau$ & $10^{-8} \ \mbox{s}$\\
bulk fluid density & $\rho$ & $1000 \mbox{ kg m}^{-3}$ \\
equilibrium surface layer density & $\rho_{Se}^s$, $\rho_{Ge}^s$ & $10^{-7} \mbox{ kg m}^{-2}$\\
bulk fluid viscosity & $\mu$ & $10^{-3}\mbox{ kg m}^{-1}\mbox{s}^{-1}$\\
coefficient of sliding friction & $\beta$ & $10^{7} \mbox{ kg m}^{-2}\mbox{s}^{-1}$\\
Darcy-type/sliding friction coefficients & $\alpha\beta$ & ${1}/{12}$ (dimensionless)$^a$
\\
entropy production constants & $U_0$, $U_{G0}$ & $10^{-4} \mbox{ m s}^{-1}$\\
equilibrium surface tension & $\sigma_{Ge}^s $ & $7\times10^{-2} \ \mbox{N m}^{-1}$\\\hline\hline
\end{tabular}
\par
\flushleft
$^a$ This value is obtained by Shikhmurzaev\cite{Shikh93} through an analogy with flow in a plane channel for the shear flow in the surface layer, and is in turn used by Billingham.\cite{Bill06,BillGrav}
\newline\bigskip
\end{minipage}

Considering $\epsilon\ll1$, we have at leading order in the bulk
\begin{equation}
\pfrac{\Bau}{\Bax} + \pfrac{\Bav}{\Baz} = 0,\qquad
\pfrac{\Bap}{\Bax}=\ppfrac{\Bau}{\Baz},\qquad
\pfrac{\Bap}{\Baz}=0,\label{eq:ndbulk}
\end{equation}
provided $\epsilon^2\Rey\ll1$, and at $z=0$ on the solid surface
\begin{equation}
\pfrac{\Bau}{\Baz} -\frac{1}{2}\pfrac{\bar \rho_S^s}{\Bax}=3\bar\beta \Bau,\qquad
\Bav = \frac{\chi_m\bar{R}_0}{3}\bar \rho_S^s,
\qquad
 \bar\tau\breve{\rho}
\left[ \frac{{\bar\alpha}}{\bar\beta}   \ppfrac{\bar\rho_S^s}{\Bax} -
\frac{3}{2}\pfrac{}{\Bax}(\Bau|_{\Baz=0})  \right] = \bar{\rho}_S^s,
\end{equation}
making the assumption $\epsilon^3\ll\lambda\breve{\rho}$. Then on the gas surface, at $z=h$, at leading order
\begin{gather}
\pfrac{\Bah}{\Bat} + \Bau \pfrac{\Bah}{\Bax} = \Bav +  \frac{\chi_m\bar{R}_0}{3} \bar \rho_G^s, \qquad
- \Bap = 3{\ppfrac{\Bah}{\Bax}}, \qquad
 -\pfrac{\Bau}{\Baz}=\pfrac{\BrG}{\Bax},
\nonumber\\
\bar{\tau}\left[\frac{1+4{\bar\alpha}}{4\bar\beta} \ppfrac{\BrG}{\Bax} - 3\pfrac{}{\Bax}{(\Bau|_{\Baz=h})}\right]
={\BrG},\label{eq:ndnormst}
\end{gather}
assuming $\epsilon^3\ll\lambda$, $\epsilon^2\bar{R}_0\ll 1$. Finally at $x=a$, $z=0$, the contact line conditions are
\begin{gather}
 \Bau - \frac{1+4{\bar\alpha}}{12\bar \beta} \pfrac{\BrG}{\Bax} -\bar{U}_{CL}+
 \breve{\rho}  \left(\frac{\Bau}{2} - \frac{{\bar\alpha}}{3\bar\beta}\pfrac{\BrS}{\Bax}- \bar{U}_{CL}\right)=0,\qquad
 \frac{2}{3}\left(\BrG +\BrS\right) = 1-\left(\pfrac{\Bah}{\Bax}\right)^2,\nonumber\\
 \breve{\rho} \left( \frac{\Bau}{2}-\frac{{\bar\alpha}}{3\bar\beta}\pfrac{\BrS}{\Bax}-  \bar{U}_{CL}\right) =  \frac{\Ca_0}{3\lambda} \left( \BrS + \BrG \right),
\end{gather}
where the same assumptions as at the solid and gas surfaces have been made, and
\begin{equation}
 \bar{U}_{CL} = \Bau - \left(\pfrac{\Bah}{\Bax}\right)^{-1}\left(\Bav + \frac{\chi_m\bar{R}_0}{3}{\BrG}\right).
\label{eq:ndcllast2}
\end{equation}
The typical values given in \refe{eq:paramsizes} support our assumptions of $\epsilon^3\ll\lambda\breve{\rho}$, $\epsilon^3\ll\lambda$, $\epsilon^2\bar{R}_0\ll 1$, and $\epsilon^2\Rey\ll1$.
The inclusion of mass transfer throughout the equations of the modern interface formation plays a role in the contact line velocity, as seen above. This can be thought of physically as being caused by the interface between the gas surface and the bulk moving at a different velocity to the actual droplet interface due to this mass transfer. In particular, we have conservation of mass from \refe{eq:cofmdim} suggesting
\begin{equation}
 \pfrac{}{t}\int_0^a h dx = \frac{\chi_m\bar{R}_0}{3} \int_0^a (\BrS+\BrG) dx. \label{eq:massconR0}
\end{equation}
For comparison, we record the form of the governing equations if Navier-slip with $\theta_d=\theta_s$ is assumed. In this case the bulk equations are as in \refe{eq:ndbulk}, and at the surfaces we would instead require
\begin{align}
 \mbox{at }\Baz=0: \quad \Bav=0, \quad \Bau = \frac{\bar\beta_{NS}}{3} \pfrac{\Bau}{\Baz} ,\quad
 \mbox{at }\Baz=\Bah: \quad \pfrac{\Bau}{\Baz} = 0, \quad -p = 3\ppfrac{\Bah}{\Bax}, \quad
 \pfrac{\Bah}{\Bat}+\Bau \pfrac{\Bah}{\Bax} = \Bav,
\end{align}
which may appear if the surface layer densities are equal to their equilibrium values, i.e. $\BrS=\BrG=0$, and $\bar\beta_{NS}=\bar\beta^{-1}$. The reduction of the conditions at the contact line for Navier-slip is less obvious, but if in addition $\breve{\rho}=0$ (so that either the fluid density at the solid surface is zero, or infinite at the gas surface) they reduce to $\Bau=\bar{U}_{CL}$ and $\theta_d=\theta_s$.

It is of interest to note the regime when the additional terms in the $\chi_m=1$ formulation are insignificant, occurring when $\bar{R}_0\ll1$. Based on the values in \refe{eq:paramsizes} this is when $1.2\times 10 ^{-2} \ll \epsilon\ll 1$, so the difference between the two interface formation models is negligible if we take $\epsilon$ in the above range.

%
%

\section{Asymptotic solution of the long-wave equations for $\chi_m=0$}
\label{sec:secchim0}

We first consider the regime where $\chi_m=0$ separately from $\chi_m=1$, as it is in this original formulation of the model where the additional contact line condition has been required in the analysis of Billingham.\cite{Bill06,BillGrav} The analysis for $\chi_m=1$ follows similarly, with details given in Sec.~\ref{sec:chim1}. The bulk equations \refe{eq:ndbulk} and the second condition in \refe{eq:ndnormst} are solved to find $\Bap =\Bap(\Bat,\Bax) = -3{\ppilfrac{\Bah}{\Bax}}$, and
\begin{equation}
 \Bau = -\frac{3\Baz^2}{2}{\pppfrac{\Bah}{\Bax}}  +  3A(\Bat,\Bax)\Baz + B(\Bat,\Bax),\qquad
 \Bav = \frac{\Baz^3}{2}{\ppppfrac{\Bah}{\Bax}}  - \frac{3\Baz^2}{2}\pfrac{A}{\Bax} - {\Baz} \pfrac{B}{\Bax} ,
\end{equation}
where we have removed an additional term arising from an integration through the use of $\Bav=0$ on the solid surface ($\Baz=0$). The quantities $3A(\Bat,\Bax)$ and $B(\Bat,\Bax)$ are the shear stress and the slip velocity on the solid surface respectively. The remaining conditions on the solid surface combine to give
\begin{equation}\label{eq:chi0AB1}
\bar\beta B - A =
\bar\tau\breve{\rho} \ppfrac{}{\Bax}
\left(
\frac{1+4{\bar\alpha}}{4}B -  \frac{\bar\alpha}{\bar\beta}A   \right).
\end{equation}
This equation in the Navier-slip model gives $B = \bar\beta_{NS} A$, so could be considered as the limit $\bar\tau\to0$. Turning attention to the gas surface equations, on $\Baz=\Bah$, the first two equations in \refe{eq:ndnormst} for $\chi_m=0$ are the same as those in the Navier-slip case leading to the free-surface equation:
\begin{equation}
 \pfrac{\Bah}{\Bat} + \pfrac{}{\Bax}\left(-
\frac{1}{2}\Bah^3 \pppfrac{\Bah}{\Bax}
+ \frac{3}{2}A\Bah^2 + B\Bah\right)=0 .\label{eq:chi0eveq}
\end{equation}
The other density conditions simplify to
\begin{equation}
\Bah\pppfrac{\Bah}{\Bax}-A =
\bar\tau\ppfrac{}{\Bax}
\left[ \frac{3}{2} \Bah^2\pppfrac{\Bah}{\Bax} - 3A\Bah - B
+ \frac{1+4{\bar\alpha}}{4\bar\beta}
\left( \Bah\pppfrac{\Bah}{\Bax}-A \right)
 \right] ,\label{eq:chi0AB2}
\end{equation}
having eliminated the $\BrG$ dependence. The equivalent Navier-slip condition is $A = \Bah \pppilfrac{\Bah}{\Bax}$, again found in the limit $\bar \tau\to0$ from the interface formation boundary condition. The above considerations for Navier-slip lead to the single equation for the motion of the free surface of
\begin{equation}
\pfrac{\Bah}{\Bat} + \pfrac{}{\Bax}\left[ \Bah^2(\Bah+\bar\beta_{NS})\pppfrac{\Bah}{\Bax} \right]=0,
\end{equation}
as expected. Finally the requirements at the contact line ($\Bax= a$, $\Baz=0$) are
\begin{gather}\nonumber
\left( 1+ \frac{\breve{\rho}}{2}\right) B
 -\bar{U}_{CL}\left( 1+ \breve{\rho}\right)-\frac{2\breve{\rho}{\bar\alpha}}{\bar\beta}(A-\bar\beta B)
= \frac{1+4{\bar\alpha}}{4\bar \beta}\left(\Bah\pppfrac{\Bah}{\Bax}-A\right)
 ,
\qquad
 1- \left(\pfrac{\Bah}{\Bax}\right)^2 = 2\mathcal{D} ,\\ \label{eq:chi0ocl3}
   \frac{B}{2}-\frac{2{\bar\alpha}}{\bar\beta}(A-\bar\beta B)- \bar{U}_{CL} =  \frac{\Ca_0}{\lambda \breve{\rho}} \mathcal{D},
\end{gather}
where $\bar{U}_{CL} = B(a)$, and here $\mathcal{D}=(\BrS+\BrG)/3$ satisfies
\begin{equation}
\mathcal{D} = {\bar\tau}\pfrac{}{\Bax} \left[
\frac{1+4{\bar\alpha}}{ 4\bar\beta} \left( \Bah\pppfrac{\Bah}{\Bax}-A
- 2{\breve{\rho}}\bar\beta B
 \right)
+
   \frac{2{\breve{\rho}}{\bar\alpha}}{\bar\beta}A    +\frac{3}{2}\Bah^2
\pppfrac{\Bah}{\Bax}  - 3A\Bah - B\right].\label{eq:chi0rhoSG}
\end{equation}
We note here that $A$ is removed in the first contact line condition when $1+4{\bar\alpha}-8\breve{\rho}{\bar\alpha}=0$, a value which is found to suggest an unphysical infinite shear stress when considering inner regions near the contact lines. It is expected that either the combination of values causing $1+4{\bar\alpha}-8\breve{\rho}{\bar\alpha}=0$ is also unphysical, or that the model assumptions break down in this case---such as the assumption that $\alpha$, $\beta$ and $\gamma$ take the same values at both solid and gas interfaces. We continue by taking parameter values away from this specific combination. Considering the contact line conditions for the Navier-slip case we take $\bar\tau\to0$ and notice that the second equation yields $\left(\pilfrac{\Bah}{\Bax}\right)^2\sim1$, suggesting that the \new{microscopic} dynamic contact angle is equal to the static contact angle ($\theta_s=\theta_d)$ at the droplet rim $\Bax=\pm a$. The remaining two contact line conditions are automatically satisfied if $\breve{\rho}=0$.

We then also require the conditions
\begin{equation}
 \Bah(a) = 0, \qquad \left.\pfrac{\Bah}{\Bax}\right|_{\Bax=0}=0, \qquad
  A(0)=B(0)=0, \qquad \int_0^a \Bah d\Bax = 1,\label{eq:starlast}
\end{equation}
corresponding respectively to zero free surface height at the droplet rim, a symmetry condition for the free surface at the droplet centre, zero shear stress and slip velocity at the centre of the wall (again by symmetry arguments), and the requirement that the cross-sectional area of the droplet remain constant.


\subsection{The outer region, $\bar\beta\gg 1$}

For convenience, we transform the domain $-a\leq\Bax\leq a$ to $-1\leq y \leq 1$ via the mapping $\Bax=ay$, and as such we are now able to write the governing equations in terms of the (non-moving) coordinate $y$, as
\begin{subequations}\label{eq:actualprob}
 \begin{align}\label{eq:eq1actualprob}
 \pfrac{\Bah}{\Bat} - \frac{\dot{a}y}{a}\pfrac{\Bah}{y} &=
 \frac{1}{a}\pfrac{}{y}\left[
\frac{\Bah^3}{2a^3} \pppfrac{\Bah}{y}
- \frac{3}{2}A\Bah^2 - B\Bah\right] ,
\\\label{eq:eq2actualprob}
\bar\beta B-A &= \frac{\bar\tau\breve{\rho}}{a^2} \ppfrac{}{y}
\left[\frac{1+4{\bar\alpha}}{4}B -  \frac{{\bar\alpha}}{\bar\beta}A
 \right],
\\
\frac{\Bah}{a^3}\pppfrac{\Bah}{y}-A &=
\frac{\bar\tau}{a^2}\ppfrac{}{y}
\left[ \frac{3\Bah^2}{2a^3}{\pppfrac{\Bah}{y}}  - 3A\Bah - B + \frac{1+4{\bar\alpha}}{4\bar\beta}
\left( \frac{\Bah}{a^3}\pppfrac{\Bah}{y}-A \right)
 \right] ,\label{eq:eq3actualprob}
\end{align}
\end{subequations}
where $\dot{a}=da/dt$ corresponds to the spreading rate of the contact line, with at the contact line $(y=1)$,
\begin{equation}
   \frac{2{\bar\alpha}}{\bar\beta}(\bar\beta B - A) - \frac{B }{2}
 = \frac{1+4{\bar\alpha}}{4\breve{\rho}\bar \beta}\left[\frac{\Bah}{a^3}\pppfrac{\Bah}{y} - A\right]
 ,
\quad
 2\mathcal{D} = 1-\left[\frac{1}{a}\pfrac{\Bah}{y}\right]^2,
\quad
\frac{2{\bar\alpha}}{\bar\beta}(\bar\beta B-A) -\frac{B}{2} =  \frac{\Ca_0}{\lambda\breve{\rho}} \mathcal{D},
\end{equation}
where
\begin{equation}
 \mathcal{D} = \frac{\bar\tau}{a}\pfrac{}{y} \left[
\frac{1+4{\bar\alpha}}{ 4\bar\beta} \left( \frac{\Bah}{a^3}\pppfrac{\Bah}{y}-A
- 2{\breve{\rho}}{\bar\beta}B
\right)
+  \frac{2{\breve{\rho}}{\bar\alpha} }{\bar\beta}A    +\frac{3\Bah^2}{2a^3}{\pppfrac{\Bah}{y}}  - 3A\Bah - B\right].\label{eq:outrgrs}
\end{equation}
Slip is not significant in the outer region, and as such we take the limit $\bar \beta\to \infinity$ \new{(with all calculations then at leading order in an asymptotic expansion in $\bar\beta^{-1}$)}, and at the same time consider the quasistatic limit $|\dot{a}|\ll1$. We introduce quasistatic expansions
of the form
\begin{subequations}\label{eq:outerexpan}
\begin{align}
 a(t) &= a_0(t) + a_1(t) + \cdots,\\
 h(x,t) &= h_0(y,a_0) + h_1(y,a_0,\dot{a}_0) + h_2(y,a_0,\dot{a}_0,a_1,\dot{a}_0^2,\ddot{a}_0) +\cdots,
 \\
 A(x,t) &= A_0(y,a_0) + A_1(y,a_0,\dot{a}_0) + A_2(y,a_0,\dot{a}_0,a_1,\dot{a}_0^2,\ddot{a}_0)+\cdots,\\
 B(x,t) &= {\bar\beta}^{-1}\left[B_0(y,a_0) + B_1(y,a_0,\dot{a}_0) + B_2(y,a_0,\dot{a}_0,a_1,\dot{a}_0^2,\ddot{a}_0)+\cdots\right],
\end{align}
\end{subequations}
where the time dependence of $h$, $A$, and $B$, enters through $a$ and its time derivatives. As written, we have assumed that corrections to the leading order radius satisfy $|a_1|\ll|\dot{a}_0|\ll a_0$, implicitly assumed by Hocking,\cite{Hocking83} and Savva and Kalliadasis.\cite{Savva09}

\new{For clarity here, we state the assumptions that will be required in our asymptotic analysis and outline the procedure to follow. There will be two main asymptotic regions. An outer region in the bulk of the droplet, where slip is not significant and an inner region at $O(1/\bar\beta)$ near the contact line (only one contact line being investigated due to the symmetry of the droplet). An intermediate region matching these two was found necessary for the Navier-slip model by Hocking,\cite{Hocking83} but here we are able to match the variables through their cubes, or negative cubes (depending on the variable at hand), with an intermediate region merely justifying this procedure. A quasistatic expansion is then performed in both inner and outer regions, and to extract the information about the spreading rate, we will find that only $a_0$ and $\dot{a}_0$ will appear. The} \new{assumptions explicitly made in the procedure are $\bar\tau=O(\bar\beta^{-1})$, $|a_1|\ll|\dot{a}_0|\ll a_0$, $\bar\beta^{-1}\ll|\dot{a}_0|$, $\ddot{a}_0=O(\dot{a}_0^2)$, and will be commented on further where they arise. The final assumption listed is merely due to the way in which the quasistatic expansions are applied, with terms involving $a_1$, $\ddot{a}_0$ and $\dot{a}_0^2$ occurring at at higher order than $O(\dot{a}_0)$, so that $a_1 = O(\ddot{a}_0) = O(\dot{a}_0^2)$. As we are only interested in the leading order terms in $\bar\beta$, all of the corrections to the radius and velocities are all greater than $O(1/\bar\beta)$ as written. This is found to be consistent, as $\dot{a}_0$ depends only logarithmically on $\bar\beta$ (as originally found by Hocking\cite{Hocking83} for Navier-slip).}

From \refe{eq:eq2actualprob} we see that $B=o(1)$, motivating the scaling of $B$ with $1 / \bar \beta$ as given in the expansion above to achieve the fullest balance. We make the further assumption that $\bar\tau\ll1$, which is reasonable based on the parameter values \refe{eq:paramsizes}. This will be confirmed after considering the inner region, where we find it necessary to have $\bar\tau=O(\bar\beta^{-1})$ to achieve $\pilfrac{h}{\Bax}|_{y=1}=O(1)$.
These considerations yield at leading order
 \begin{equation}
A_0 = B_0 = \frac{\Bah_0}{a_0^3} \pppfrac{\Bah_0}{y},\qquad \pfrac{}{y}\left( \Bah_0^3\pppfrac{\Bah_0}{y} \right)=0,\label{eq:outle}
\end{equation}
with conditions
\begin{equation}
A_0(1)=0,\qquad \left.\left(\pfrac{\Bah_0}{y}\right)^2\right|_{y=1}=a_0^2,\qquad
  \Bah_0(1) = 0, \qquad \left.\pfrac{\Bah_0}{y}\right|_{y=0}=0,
\qquad \int_0^1 \Bah_0 dy = \frac{1}{a_0}. \label{eq:usualdropbcs}
\end{equation}
Equations \refe{eq:outle} and conditions \refe{eq:usualdropbcs} \new{(apart from the contact angle condition)} are solved to find
\begin{equation}
 \Bah_0 = \frac{3(1-y^2)}{2a_0}
,\qquad \mbox{giving} \qquad A_0 = B_0 =0. \label{eq:H0sol}
\end{equation}
The equations for the next term are
 \begin{equation}
\left.
\begin{array}{l}
\displaystyle A_1 = B_1 
=  \frac{\Bah_0}{a_0^3} 
 \pppfrac{\Bah_1}{y}
\\
 \displaystyle \pppfrac{\Bah_1}{y}
=  \frac{4\dot{a}_0 a_0^5 y}{ 9(1-y^2)^2} 
\end{array}
\right\},
\quad
\mbox{with solutions}
\quad \left.
\begin{array}{l}
\displaystyle
A_1 = B_1 = \frac{2\dot{a}_0 a_0 y}{ 3(1-y^2)}\\
\displaystyle
\pfrac{\Bah_1}{y} = \frac{1}{9}\dot{a}_0a_0^5\left[\ln\left( \frac{1+y}{1-y} \right) - 3y\right]
\end{array}\right\},\label{eq:outH1}
\end{equation}
having used $\pilfrac{\Bah_0}{\Bat}=\dot{a}_0\pilfrac{\Bah_0}{a_0}$, and where arbitrary constants have been determined from $h_1(1)=0$ and $(\pilfrac{\Bah_1}{y})(0)=0$. The solutions for $A_1$ and $B_1$ automatically satisfy the symmetry conditions $A_1(0)=B_1(0)=0$. The behaviours of the free surface slope, $A_1$ and $B_1$ as we approach the contact line at this order are then found as
\begin{equation}
 -\pfrac{\Bah}{\Bax} \sim \frac{3}{a_0^2} + \frac{1}{9}{\dot{a}_0a_0^4}\ln\left[\frac{\mye^3(a_0-\Bax)}{2a_0}\right],\qquad
 A \sim \frac{1}{3}\dot{a}_0a_0^2(a_0-\Bax)^{-1}, \qquad
 B \sim \frac{1}{3\bar\beta}\dot{a}_0a_0^2(a_0-\Bax)^{-1}.\label{eq:ABfirstbehav}
\end{equation}
as $\Bax\to a$. The behaviours of $A$ and $B$ in \refe{eq:ABfirstbehav} include only one term in the expansion for $|\dot{a}|\ll1$ as the $O(1)$ terms are zero. As is seen in \new{a number of other works with slip\cite{Hocking83,Savva09} and precursor films\cite{PismenEggersSolv}} for the matching of $\pilfrac{\Bah}{y}$, the logarithmic terms appearing in the second term must be included for the correct matching procedure. We expect the same to be necessary for the matching of $A$ and $B$, so we continue to $O(|\dot{a}_0|^2)$ to find terms of that size. Returning to \refe{eq:actualprob} with expansions \refe{eq:outerexpan}, we have
\begin{subequations}
\begin{align}
0&={a_0^4}\left(\dot{a}_0\pfrac{\Bah_1}{a_0}+\ddot{a}_0\pfrac{\Bah_1}{\dot{a}_0}\right)- {a_0^3}{\dot{a}_0y}\pfrac{\Bah_1}{y}
 + \pfrac{}{y}\left(
\Bah_0^3\pppfrac{\Bah_2}{y} + 3\Bah_0^2\Bah_1\pppfrac{\Bah_1}{y} \right),
\\
A_2 = B_2 &= \frac{1}{a_0^3}\left( \Bah_0 \pppfrac{\Bah_2}{y} + \Bah_1 \pppfrac{\Bah_1}{y}\right),
\end{align}
\end{subequations}
having used \refe{eq:outH1}. We may have expected terms involving the correction to the radius $a_1$ in the above expressions, but it is readily seen that the terms arising are $O(a_1\dot{a}_0)$, and hence do not feature at this order. Using \refe{eq:H0sol}, \refe{eq:outH1} and applying $\int_0^1 \Bah_1 dy = 0$, these may be solved to determine the behaviours
\begin{equation}
A_2 = B_2 \sim -\frac{a_0^7 \dot{a}_0^2}{3^4(1-y)} \ln \left[\frac{\mye^2(1-y)}{2} \right], \qquad \mbox{as }y\to1,
\end{equation}
noting that there are no terms involving $\ddot{a}_0$ at this order. In the expansions for $A_2$ and $B_2$, they occur first at $O(\ddot{a}_0\ln(1-y))$, and are smaller than the terms kept due to the assumption that $\ddot{a}_0=O(\dot{a}_0^2)$. Collecting all the results together, our two term matching conditions become
\begin{gather}\nonumber
 -\pfrac{\Bah}{\Bax} \sim \frac{3}{a_0^2} + \frac{1}{9}{\dot{a}_0a_0^4}\ln\left[\frac{\mye^3(a_0-\Bax)}{2a_0}\right],\\
 A \sim \frac{1}{3}\frac{\dot{a}_0a_0^2}{a_0-\Bax}-\frac{1}{3^4}\frac{a_0^8 \dot{a}_0^2}{a_0-\Bax} \ln \left[\frac{\mye^2(a_0-\Bax)}{2a_0} \right], \qquad
 B \sim \frac{1}{3\bar\beta}\frac{\dot{a}_0a_0^2}{a_0-\Bax}-\frac{1}{3^4\bar\beta}\frac{a_0^8 \dot{a}_0^2}{a_0-\Bax} \ln \left[\frac{\mye^2(a_0-\Bax)}{2a_0} \right],\label{eq:outmatch}
\end{gather}
as $\Bax\to a$. These asymptotic expansions will be matched to their counterparts from an inner region close to the contact line to determine the droplet spreading rate. The matching is carried out in a region such that for $A$ and $B$ the $O(\dot{a}_0)$ terms occur at higher order than the $O(\dot{a}_0^2\ln(a_0-\Bax)(a_0-\Bax)^{-1})$ terms and the $O(\dot{a}_0(a_0-\Bax)^{-1})$ terms respectively, and equivalent restrictions for $\pilfrac{h}{x}$, to ensure asymptotic validity.


\subsection{$O(1/\bar\beta)$ Inner regions}

The outer solution found does not satisfy the boundary conditions at the contact line, and as such motivates the presence of a boundary layer, or inner region, close to the rim of the droplet. In fact, for the slope of the free surface defining the contact angle to satisfy the boundary conditions, we would require $a^2 = 3$. If the initial droplet radius satisfies this value, the droplet will remain in equilibrium, but radii different from this will cause motion of the free surface to return to this equilibrium. Considering distances $a-\Bax=O(1/\bar{\beta})$, we make the change of variables
\begin{equation}
 \Bax = a - {\xi}{\bar{\beta}}^{-1},\quad \Bah = {\Psi}{\bar{\beta}}^{-1}, \quad A = \bar{\beta}\hat{A}, \quad B=\hat{B},
\label{eq:innerscals}
\end{equation}
where the scalings for $\Bah$, $A$ and $B$ are motivated by their forms in the outer solution. We then find the governing equations become
\begin{subequations}\label{eq:inner}
\begin{align}\label{eq:inner1}
\frac{1}{\bar{\beta}}\pfrac{\Psi}{\Bat}+\dot{a}\pfrac{\Psi}{\xi} &= \pfrac{}{\xi}\left[\frac{1}{2}\Psi^3 \pppfrac{\Psi}{\xi}
+ \frac{3}{2}\hat{A}\Psi^2 + {\hat{B}}\Psi\right] ,\\ \label{eq:inner2}
\hat A-\hat B &= \bar\beta \bar\tau\breve{\rho}
\ppfrac{}{\xi}
\left[  \bar\alpha\hat A - \frac{1+4{\bar\alpha}}{4}\hat B   \right],\\
\Psi\pppfrac{\Psi}{\xi}+\hat A &=
\bar\beta\bar\tau
\ppfrac{}{\xi}
\left[ \frac{3}{2}\Psi^2{\pppfrac{\Psi}{\xi}}  + 3\hat A\Psi + \hat B
+ \frac{1+4{\bar\alpha}}{4} \left(\Psi\pppfrac{\Psi}{\xi}+\hat A \right)
 \right] ,\label{eq:inner3}
\end{align}
\end{subequations}
with, at $\xi=0$, the contact line conditions
\begin{equation}
 \hat{B} + 4 \bar\alpha(\hat A-\hat B)
 = \frac{1+4{\bar\alpha}}{2\breve{\rho}}\left[ \Psi\pppfrac{\Psi}{\xi}+\hat A\right]
 ,
\qquad
2\mathcal{D} = 1-\left(\pfrac{\Psi}{\xi}\right)^2,
\qquad
 4{\bar\alpha}(\hat B - \hat A) -\hat B = \frac{2\Ca_0}{\lambda\breve{\rho}} \mathcal{D},
\end{equation}
where
\begin{equation}
 \mathcal{D} = {\bar\beta\bar\tau}\pfrac{}{\xi}  \left[
\frac{1+4{\bar\alpha}}{4} \left( \Psi\pppfrac{\Psi}{\xi}+\hat A
+ 2{\breve{\rho}}\hat B
\right)
- 2{\breve{\rho}} {\bar\alpha} \hat A  + \frac{3\Psi^2}{2}{\pppfrac{\Psi}{\xi}}  + 3\hat A\Psi + \hat B\right].\label{eq:chi0inrhoGS}
\end{equation}
We then also require $\Psi(0) = 0$ and that the solution matches to the outer region. The dependence on the combined parameter $\bar{\beta}\bar{\tau}$ in the above equations is clear, and that to construct an asymptotic solution with the contact angle $O(1)$ requires $\bar\beta\bar\tau=O(1)$ as assumed in the outer region. Noting that $\bar{\beta}\bar{\tau}=O(2\times10^3\epsilon^2)$ from \refe{eq:paramsizes} for water at room temperature, the above assumption is a reasonable one, but we will consider a range of values. It is noteworthy that $\bar{\beta}\bar{\tau}$ was also found to be an important parameter in the gravity driven thin-film flow situation.\cite{BillGrav} As seen for the Navier-slip model,\cite{Savva09} and also in the comparison between slip and precursor film models,\cite{SavvaPrecursorSlip} the inner region solution as $\xi\to\infinity$ will determine the matching condition to the outer region and provide the relationship between the spreading rate $\dot{a}$ and the radius of the droplet $a$. We will proceed by considering a quasistatic expansion for the inner equations, but first we note that there is a correspondence between the limit $\xi\to\infinity$ and the situation when $\bar\beta\bar\tau\ll1$. In particular, given that $\Psi$, $\hat{A}$ and $\hat{B}$ cannot display exponential growth as $\xi\to\infinity$, and \new{also that for matching to the outer region solutions \refe{eq:outmatch}} we require $\Psi/\xi^2\to0$, $\hat{A}\to0$, and $\hat{B}\to0$ as $\xi\to\infinity$, then we have to leading order
\begin{equation}\label{eq:fullinmatch}
 \hat{A} \sim \hat{B} \sim -\Psi\pppfrac{\Psi}{\xi},\qquad
 \frac{1}{\bar{\beta}}\pfrac{\Psi}{\Bat}+
\dot{a}\pfrac{\Psi}{\xi} \sim -\pfrac{}{\xi}\left[\Psi^2(1+\Psi)\pppfrac{\Psi}{\xi} \right] , \qquad \mbox{as }\xi \to \infinity.
\end{equation}

\subsubsection{The inner region with quasistatic spreading}

To progress with the analysis of the inner region governing equations \refe{eq:inner}, we assert that $\hat{A}=O(\dot{a}_0)$, $\hat{B}=O(\dot{a}_0)$. This is motivated by observing that ${A}=O(\dot{a}_0)$, ${B}=O(\dot{a}_0)$ in the outer solution, thus for solutions of $\hat{A}$ and $\hat{B}$ in the inner region to match, they must be at most $O(\dot{a}_0)$. Alternatively, the original variables $A$ and $B$ correspond to the shear stress and velocity at the solid surface so would not be expected to contribute to the leading order static terms. Introducing the quasistatic expansions
\begin{subequations}
\begin{align}
 a(t) &= a_0(t) + a_1(t) + \cdots,\\
 \Psi(\xi, \Bat) &= \Psi_0(\xi,a_0) + \Psi_1(\xi,\dot{a}_0,a_0) + \Psi_2(\xi,a_0,\dot{a}_0,a_1,\dot{a}_0^2,\ddot{a}_0) + \cdots, \\
 \hat A(x,t) &= \hat A_1(\xi,\dot{a}_0,a_0) + \hat A_2(\xi,a_0,\dot{a}_0,\ldots) + \cdots,\\
 \hat B(x,t) &= \hat B_1(\xi,\dot{a}_0,a_0) + \hat B_2(\xi,a_0,\dot{a}_0,\ldots) + \cdots,
\end{align}\label{eq:inquas}
\end{subequations}
then \refe{eq:inner}--\refe{eq:chi0inrhoGS} at leading order are
\begin{equation}\label{eq:inquasadot}
 0=\pfrac{}{\xi}\left(\frac{1}{2}\Psi_0^3 \pppfrac{\Psi_0}{\xi}\right) ,\qquad
\Psi_0\pppfrac{\Psi_0}{\xi} =
\bar\beta\bar\tau\ppfrac{}{\xi}
\left[ \Psi_0\pppfrac{\Psi_0}{\xi}\left(\frac{3}{2}\Psi_0 + \frac{1+4{\bar\alpha}}{4}  \right)\right],
\end{equation}
with the contact line conditions
\begin{gather}
\Psi_0\pppfrac{\Psi_0}{\xi}=0, \qquad
2\bar\beta\bar\tau\pfrac{}{\xi} \left[\Psi_0\pppfrac{\Psi_0}{\xi}\left(\frac{3}{2}\Psi_0
+ \frac{1+4{\bar\alpha}}{4}  \right)
 \right]
 = 1- \left(\pfrac{\Psi_0}{\xi}\right)^2,
\nonumber\\
\pfrac{}{\xi} \left[\Psi_0\pppfrac{\Psi_0}{\xi}\left(\frac{3}{2}\Psi_0 + \frac{1+4{\bar\alpha}}{4}  \right)
 \right]=0,\label{eq:allbcABadot}
\end{gather}
at $\xi=0$. As $\Psi_0(0)=0$, we find the first equation of \refe{eq:inquasadot} gives $ \Psi_0 = K_0 \xi^2 + K_1 \xi$, which automatically satisfies the second equation and the first and third contact line conditions. Requiring $\Psi_0/\xi^2\to0$ as $\xi\to\infinity$ to match to the outer region and applying the second contact line condition gives $\Psi_0 = \xi$. We note that the third contact line condition in \refe{eq:allbcABadot} is that added by Bedeaux\cite{BedeauxExtraCond} and Billingham,\cite{Bill06,BillGrav} and whilst it is automatically satisfied here, it was not required to determine the solution at this order.

Having determined the leading order free surface behaviour in the inner region, we consider the next order in the governing equations, where they satisfy
\begin{subequations}\label{eq:inneradot}
\begin{align}\label{eq:inneradot1}
\dot{a}_0 &= \pfrac{}{\xi}\left[\frac{1}{2}\xi^3 \pppfrac{\Psi_1}{\xi}
+ \frac{3}{2}\hat{A}_1\xi^2 + {\hat{B}_1}\xi\right] ,\\ \label{eq:inneradot2}
\hat A_1-\hat B_1 &= \bar\beta \bar\tau\breve{\rho}
\ppfrac{}{\xi}
\left[ \bar\alpha\hat A_1 - \frac{1+4{\bar\alpha}}{4}\hat B_1 \right],\\
\xi\pppfrac{\Psi_1}{\xi}+\hat A_1 &=
\bar\beta\bar\tau\ppfrac{}{\xi}
\left[ \frac{3}{2}\xi^2{\pppfrac{\Psi_1}{\xi}} + 3\hat A_1\xi + \hat B_1
+ \frac{1+4{\bar\alpha}}{4}
\left(\xi\pppfrac{\Psi_1}{\xi}+\hat A_1 \right)
 \right] ,\label{eq:inneradot3}
\end{align}
\end{subequations}
having neglected the $\bar{\beta}^{-1}\pilfrac{}{\Bat}\Psi_0 = {\dot{a}_0\xi}(\bar\beta a_0)^{-1}$ term as $|\dot{a}_0|\ll\bar{\beta}|\dot{a}_0|$. At $\xi=0$, we have
\begin{equation}
\hat{B}_1 +
 4  {{\bar\alpha}}(\hat A_1-\hat B_1)
 = \frac{1+4{\bar\alpha}}{2\breve{\rho}}\left[ \xi\pppfrac{\Psi_1}{\xi}+\hat A_1\right]
 ,
\qquad
-\mathcal{D} = \pfrac{\Psi_1}{\xi},
\qquad
   4\bar\alpha(\hat B_1 - \hat A_1)-\hat B_1   = \frac{2\Ca_0}{\lambda\breve{\rho}} \mathcal{D},
\end{equation}
where
\begin{align}
 \mathcal{D} = \bar\beta\bar\tau\pfrac{}{\xi} & \left[
\frac{1+4{\bar\alpha}}{4} \left( \xi\pppfrac{\Psi_1}{\xi}+\hat A_1
+ 2{\breve{\rho}}\hat B_1
\right)
-2 {\breve{\rho}} {\bar\alpha} \hat A_1  + \frac{3\xi^2}{2}{\pppfrac{\Psi_1}{\xi}}  + 3\hat A_1\xi + \hat B_1\right],\label{eq:chi0inOadotrhoGS}
\end{align}
with $\Psi_1(0)=0$. We also require $\Psi_1/\xi^2\to0$, $\hat{A}_1\to0$, and $\hat{B}_1\to0$ as $\xi\to\infinity$ to match. As noted previously, it is the behaviour as $\xi\to\infinity$ \new{that is} crucial to the leading order spreading dynamics of the droplet, so we proceed by considering an asymptotic expansion of the form
\begin{equation}
 \frac{\Psi_1}{\dot{a}_0} \sim  \sum_{i=0}^\infinity \xi^{1-i}\left(c_{i1}  \ln\xi + c_{i2}  \right),
\end{equation}
with similar expressions for $\hat A_1$, $\hat B_1$, to find
\begin{align}\label{eq:A1B1xi2inf}
 \Psi_1 = \dot{a}_0 \left[\xi\ln\xi + c_{0} \xi + O(\ln\xi)\right],
 \qquad
 \hat A_1 = \dot{a}_0 \left[\xi^{-1} + O(\xi^{-2}) \right],
 \qquad
 \hat B_1 = \dot{a}_0 \left[\xi^{-1} + O(\xi^{-2}) \right],
\end{align}
as $\xi\to\infinity$, in agreement with \refe{eq:fullinmatch} at $O(\dot{a}_0)$. The constant $c_0$ is unknown here---it will depend on all of the parameters of the problem, and must be determined from the full solution of \refe{eq:inneradot}--\refe{eq:chi0inOadotrhoGS}, considered in Sec.~\ref{sec:fullOadotchi0}. The required two terms have been found for the matching of the slope in the inner region. As in the outer region, we require terms in both $\dot{a}_0$ and $\dot{a}_0^2$ (and possibly other terms at this order) for $\hat{A}$ and $\hat{B}$ to also have a two term expansion. From \refe{eq:fullinmatch}, we have that
\begin{align}
 \hat{A}_2 \sim \hat{B}_2 \sim -\xi\pppfrac{\Psi_2}{\xi} - \Psi_1\pppfrac{\Psi_1}{\xi},
\qquad
 \dot{a}_0\pfrac{\Psi_1}{\xi} \sim - \pfrac{}{\xi}\left( \xi^3\pppfrac{\Psi_2}{\xi} + 3\xi^2\Psi_1\pppfrac{\Psi_1}{\xi} \right),
\end{align}
as $\xi\to \infinity$, where we have neglected the terms including $\pilfrac{\Psi}{t}$ by making the assumption $\bar\beta^{-1}\ll|\dot{a}_0|$, an assumption also made by Hocking.\cite{Hocking83} This assumption means that our asymptotic analysis is not valid very close to equilibrium, although we will see from our comparison with numerical results for the full problem that excellent agreement is still found. Using \refe{eq:A1B1xi2inf} and combining all of these results in terms of outer variables
\begin{equation}\label{eq:matchfrominner}
-\pfrac{\Bah}{\Bax} \sim 1 + \dot{a}_0\ln\left[\mye^{1 + c_0} \bar{\beta}(a_0-\Bax) \right], \qquad
 {A} \sim \bar\beta B \sim
\frac{\dot{a}_0}{a_0-\Bax} - \frac{\dot{a}_0^2}{a_0-\Bax} \ln[\mye^{c_0}\bar{\beta}(a_0-\Bax)] ,
\end{equation}
as $\bar{\beta}(a-\Bax)\to\infinity$, which must be matched to \refe{eq:outmatch}. It is apparent, as for the Navier-slip situation, that the logarithmic terms of the free surface slope (and of $A$ and $B$) in the inner and outer regions do not generally match. This may be resolved by considering an intermediate region as was done first by Hocking,\cite{Hocking83} however Savva and Kalliadasis noted that matching may be resolved by considering $\left(\partial_{\Bax}{\Bah} \right)^3$ between inner and outer regions, with the intermediate region simply justifying this step.\cite{Savva09} Equivalently, here an intermediate region motivated by the analysis of Hocking merely justifies the matching of $(\pilfrac{\Bah}{\Bax})^3$, $A^{-3}$ and $B^{-3}$.

\begin{figure}[ht]
    \centering
    \includegraphics{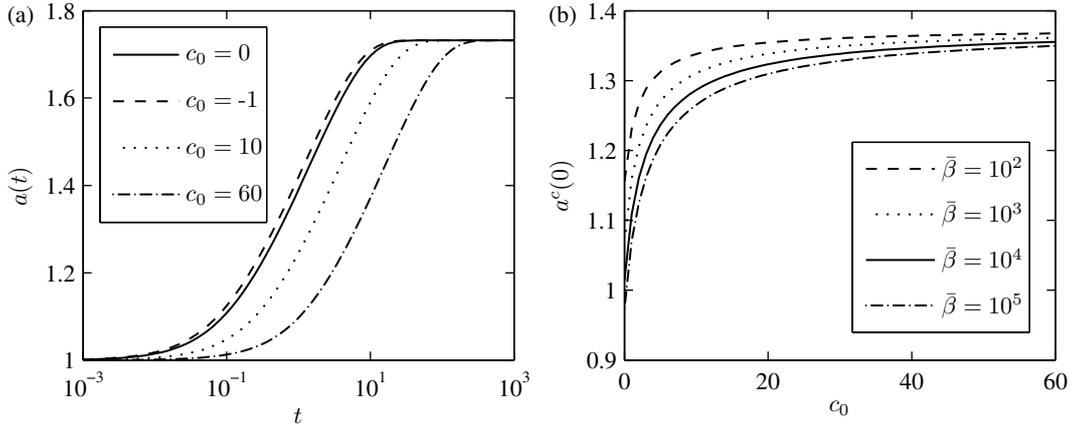}
    \caption{(a) Evolution of the droplet radius $a(\Bat)$ for various values of $c_0$ from initial radius $a(0)=1$ and $\bar\beta = 10^3$ showing delayed spreading for increasing $c_0$. (b) The critical initial value of the droplet radius $a^c(0)$ for varying $c_0$ and $\bar\beta$. Asymptotic validity if $a(0)$ is closer to the equilibrium radius, $a(\infinity)=\sqrt{3}$, than the value of $a^c(0)$ here.}
    \label{fig:asymvalid}
\end{figure}

A comparison of the cube of the free surface slope and the negative cube of $A$ and $B$ from the outer expansions \refe{eq:outmatch} and the inner expansions \refe{eq:matchfrominner} then gives the relationship between the spreading rate $\dot{a}_0$ and the droplet radius $a_0$ as
\begin{equation}
 3\dot{a}_0a_0^6\ln\left( \frac{\mye^{2-c_0}}{2a_0\bar\beta} \right) = a_0^6-27. \label{eq:adotina}
\end{equation}
This spreading rate has a similar form to the result for Navier-slip, where $\beta_{NS}=1/\bar\beta$ and $c_0=0$. This means that the interface formation model (with $\chi_m=0$) has a spreading rate equivalent to a Navier-slip model (with $\theta_d$=$\theta_s$) with slip length $\beta_{NS}=\mye^{-c_0}/\bar\beta$, although we will also see an equivalence with other slip models with velocity dependent \new{microscopic} dynamic contact angles. From \refe{eq:adotina} it is evident that provided $a_0$ is away from equilibrium, $a(\infinity)=\sqrt{3}$, then we have $a_0=(ct)^{1/7}$, where $c=-63/\ln[ \mye^{2-c_0}/(2a_0\bar\beta) ]$. Neglecting the weak logarithmic dependence of time on $c$, this gives agreement with the spreading of Tanner's law.\cite{TannersLaw, mchale_stripes} As the equilibrium radius is approached the behaviour becomes exponential, so that $a_0\sim a_\infinity - \bar\epsilon \mye^{-w t}$, where $\bar\epsilon\ll1$ is some constant and
\begin{equation}
 w = -\frac{2}{\sqrt{3}}\left[\ln\left( \frac{\mye^{2-c_0}}{2\sqrt{3}\bar\beta} \right)\right]^{-1}.
\end{equation}
These spreading behaviours of the droplet radius both reduce to equivalent expressions for Navier-slip when $c_0=0$, as expected (see Savva and Kalliadasis,\cite{Savva09} taking the planar substrate situation). Figure \ref{fig:asymvalid}(a) shows the evolution of the droplet radius $a(\Bat)$ for various values of $c_0$ from initial radius $a(0)=1$ and $\bar\beta = 10^3$ using the spreading rate found in \refe{eq:adotina}, showing that increasing $c_0$ delays the spreading process, but that the final equilibrium radius is unaffected.

The asymptotic expansions \refe{eq:matchfrominner} are valid for
\begin{equation}
 \Bax \gg a_0 - \frac{1}{\bar\beta}\exp\left( \frac{1}{|\dot{a}_0|}-1-c_0 \right),
\end{equation}
and for matching between inner and outer regions we have $\xi = \bar\beta(a-\Bax) \gg 1$. Together these require
\begin{equation}
 (1+c_0)|\dot{a}_0|\ll1.\label{eq:c0adotvalid}
\end{equation}
This reduces to the familiar requirement $|\dot{a}_0|\ll1$ for Navier-slip, $c_0=0$, but imposes a greater restriction on the spreading rate for positive $c_0$. Using \refe{eq:adotina}, we show the critical value of the initial radius $a^c(0)$ in Fig.~\ref{fig:asymvalid}(b), so that when droplet spreading starts closer to equilibrium than that value, the asymptotic procedure is valid. As noted above, we also required $\bar\beta^{-1}\ll|\dot{a}_0|$ for the asymptotic analysis to hold. Once the droplet has spread sufficiently for this to be invalidated, the droplet will already be negligibly close to equilibrium, as our analysis is based on the assumption that $\bar\beta\gg1$. \new{We note that an asymptotic investigation to relax the quasistatic assumption has been performed by Eggers for a slip model in the vicinity of a contact line.\cite{eggershigherCa} Even for this simpler model, matching to an outer flow region (such as required for droplet spreading) was deemed a challenging and open problem, and thus it is also beyond the scope of the present work with the interface formation model.}

%
%

\subsubsection{The full $O(\dot{a}_0)$ inner region behaviour}
\label{sec:fullOadotchi0}

We drop the subscript $0$ from $a_0$ and $\dot{a}_0$ for the remainder of this section as we have found that corrections to the radius do not enter into our analysis. Returning to the $O(\dot{a})$ inner region equations \refe{eq:inneradot}--\refe{eq:chi0inOadotrhoGS}, we now wish to find a full solution to complete our asymptotic description and obtain an $O(\dot{a})$ correction to the \new{microscopic} contact angle. Equation \refe{eq:inneradot1} may be solved to give
\begin{equation}
 \hat{B}_1 = \dot{a} - \frac{1}{2} \xi^2 \pppfrac{\Psi_1}{\xi} - \frac{3}{2}\hat{A}_1\xi, \label{eq:B1sol}
\end{equation}
having used $\Psi_1\to 0$, $\hat{A}_1\xi\to0$, $\hat{B}_1\xi\to0$ as $\xi\to0$, to remove a constant of integration---\new{the latter two conditions being imposed as the shear stress and velocity (or correspondingly $\hat{A}$ and $\hat{B}$) at the solid surface should at worst be logarithmically singular as the contact line is approached (they will, in fact, be found to be finite in later calculations)}. This determines $\hat{B}_1(0)=\dot{a}$, and reduces the problem to a seventh order system of two ODEs
\begin{subequations}\label{eq:inadode}
\begin{align}
( 2 + 3\xi )\hat A_1 +\xi^2 \pppfrac{\Psi_1}{\xi}  &=  2\dot{a}
  + \frac{\bar\beta \bar\tau\breve{\rho}}{4}
\ppfrac{}{\xi}
\left[
(1+4{\bar\alpha}) \xi^2 \pppfrac{\Psi_1}{\xi}
+
\left( \frac{3\xi}{1+4{\bar\alpha}} + 8{\bar\alpha} \right)\hat A_1
\right],
\\
\xi\pppfrac{\Psi_1}{\xi}+\hat A_1 &=
\frac{\bar\beta\bar\tau}{4}\ppfrac{}{\xi}
\left[ \left(4\xi+{1+4{\bar\alpha}}\right)\xi\pppfrac{\Psi_1}{\xi}
+ \left({6\xi}+{1+4{\bar\alpha}} \right)\hat A_1
 \right] ,
\end{align}
\end{subequations}
subject to at $\xi=0$
\begin{subequations}
\begin{align}
\label{eq:inneradotcl1}
\frac{1+4{\bar\alpha}-8\breve{\rho}{\bar\alpha}}{1+4{\bar\alpha}}\hat{A}_1
+\xi\pppfrac{\Psi_1}{\xi} &= \frac{2\breve{\rho}(1-4{\bar\alpha})}{1+4{\bar\alpha}}\dot{a}
,\\\label{eq:inneradotcl2}
-\mathcal{D} &= \pfrac{\Psi_1}{\xi}
,\\
(1-4{\bar\alpha})\dot{a}+4\bar\alpha\hat{A}_1  &=  -\frac{2\Ca_0}{\lambda \breve{\rho}} \mathcal{D},
\label{eq:inneradotcl4}
\end{align}
\end{subequations}
with $\Psi_1 =0$, and where
\begin{equation}
 \mathcal{D} = \frac{\bar\beta\bar\tau}{4}\pfrac{}{\xi} \left[(
    6\xi + (1+4{\bar\alpha})(1 - 3\xi\breve{\rho}) -8\breve{\rho}{\bar\alpha}) \hat A_1
+  (4\xi + (1+4{\bar\alpha})(1 -\xi\breve{\rho}) )\xi \pppfrac{\Psi_1}{\xi}\right].
\end{equation}
We note that the boundary conditions are able to support a solution with $(\xi\partial^3_\xi \Psi_1)(0)=0$, which allows the model to alleviate the logarithmic pressure singularity associated with the Navier-slip model. \new{This weak singularity occurs at the contact line for Navier-slip (seen for example, in Refs.~\onlinecite{ShikhSingualarities06,SprittlesShikhCornerNumArt}), where a finite force is nevertheless predicted. We will return to this issue when considering the asymptotic behaviour of the interface formation model as we approach the contact line, where comparisons between models will be clearer.} The ODE system \refe{eq:inadode} is subject to the contact line conditions at the singular point $\xi=0$ and the matching conditions as $\xi\to\infinity$ of $\hat{A}_1\to0$, $\hat{B}_1\to0$, $\Psi_1/\xi^2\to0$, all \new{being} satisfied by behaviour \refe{eq:A1B1xi2inf}. We perform a local analysis about these points separately to determine the relevant asymptotic behaviours.

\paragraph*{Eigenmode analysis:}
We are interested in imposing the contact line and matching behaviours
\begin{equation}
 \Psi_1 \sim \hat{P}_0\xi, \quad\hat{A}_1 \sim \hat{S}_0, \quad \mbox{as } \xi\to0, \qquad \mbox{and} \qquad
 \Psi_1 \sim \dot{a} \left[\ln\xi+c_0\right]\xi, \quad \hat{A}_1\sim \dot{a}{\xi}^{-1}, \quad \mbox{as } \xi\to\infinity,
\end{equation}
on the ODE system \refe{eq:inadode}, where $\hat{P}_0$ and $\hat{S}_0$ are the coefficients of the leading order terms in the expansions as $\xi\to0$ of $\Psi_1$ and $\hat{A}_1$ respectively. To determine the correct specification of the inner problem the degrees of freedom contained within the asymptotic behaviours must be found, determining the number of boundary conditions imposed. We perform an eigenmode analysis for the behaviour as $\xi\to\infinity$ by considering
\begin{equation}
 \Psi_1 = \dot{a} \left[\ln\xi+c_0\right]\xi + \bar{\delta}\bar{\Psi}_1(\xi), \quad
 \hat{A}_1 = {\dot{a}}{\xi}^{-1} + \bar{\delta}\bar{A}_1(\xi)
\end{equation}
where $\bar\delta$ is a small artificial gauge. Keeping terms of $O(\bar\delta)$, the resulting equations in $\bar{\Psi}_1,\bar{A}_1$ have the seven asymptotic behaviours for the eigenmodes of
\begin{gather}
\left.\begin{array}{l}\bar{\Psi}_1 = 1\\ \bar{A}_1 = 0\end{array}\hspace{-4pt}\right\},
 \left.\begin{array}{l}\bar{\Psi}_1 = \xi\\ \bar{A}_1 = 0\end{array}\hspace{-4pt}\right\} ,
 \left.\begin{array}{l}\bar{\Psi}_1 = \xi^2\\ \bar{A}_1 = 0\end{array}\hspace{-4pt}\right\},
\nonumber\\
\left.\begin{array}{l}\bar{\Psi}_1 \sim \exp\left[ \pm \frac{2\xi}{\sqrt{\breve{\rho}\bar\beta\bar\tau(1+4{\bar\alpha})}} \right]
\\ \bar{A}_1 \sim \mp \frac{2}{3}\left[\frac{4}{\breve{\rho}\bar\beta\bar\tau(1+4{\bar\alpha})}\right]^{3/2}\xi
\exp\left[ \pm \frac{2\xi}{\sqrt{\breve{\rho}\bar\beta\bar\tau(1+4{\bar\alpha})}} \right]
\end{array}\hspace{-4pt}\right\} ,
\left.\begin{array}{l}\bar{\Psi}_1 \sim \exp\left[ \pm \frac{4\sqrt{\xi}}{\sqrt{3\bar\beta\bar\tau}} \right]
\\ \bar{A}_1 \sim \mp \frac{8\sqrt{3}}{(9\bar\beta\bar\tau)^{3/2}}\frac{1}{\sqrt\xi}
\exp\left[ \pm \frac{4\sqrt{\xi}}{\sqrt{3\bar\beta\bar\tau}} \right]\label{eq:ffmodes}
\end{array}\hspace{-4pt}\right\}.
\end{gather}
The second mode corresponds to changes in the parameter $c_0$, which has to be determined numerically, with the first mode corresponding to another degree of freedom at higher order in the asymptotic expansion. The third mode and the two positive exponential modes are inconsistent with the behaviour as $\xi\to\infinity$, and as such impose three conditions on \refe{eq:inadode}. The two negative exponential modes are exponentially small as $\xi\to\infinity$ and correspond to two further degrees of freedom. This implies that the contact line expansions are not analytic, with the presence of exponentially small terms. These expansions take the forms 
\begin{subequations}\label{eq:psi1simxiinf}
\begin{align}
 \Psi_1 &\sim \dot{a}\left[ \left(\ln\xi+c_0\right)\xi + \left(\frac{1}{2}\ln\xi+c_1\right) + \frac{3\bar\beta\bar\tau+2}{12}\xi^{-1} \right] 
+ \cdots
+ c_2\exp\left[ - \frac{4\sqrt{\xi}}{\sqrt{3\bar\beta\bar\tau}} \right]
\nonumber\\  &
+ \cdots
+ c_3\exp\left[ - \frac{2\xi}{\sqrt{\breve{\rho}\bar\beta\bar\tau(1+4{\bar\alpha})}} \right]
+ \cdots ,
\\
\hat{A}_1 &\sim \dot{a}\left[ \frac{1}{\xi} - \frac{1}{\xi^2} + \frac{\bar\beta\bar\tau+2}{2\xi^{3}} \right] + \cdots
+ \frac{8c_2\sqrt{3}}{(9\bar\beta\bar\tau)^{3/2}}\frac{1}{\sqrt\xi}
\exp\left[ - \frac{4\sqrt{\xi}}{\sqrt{3\bar\beta\bar\tau}} \right]  + \cdots
\nonumber\\ &
+ \frac{2c_3\xi}{3}\left[\frac{4}{\breve{\rho}\bar\beta\bar\tau(1+4{\bar\alpha})}\right] ^{3/2}
\exp\left[ -\frac{2\xi}{\sqrt{\breve{\rho}\bar\beta\bar\tau(1+4{\bar\alpha})}} \right]
+ \cdots
,\end{align}
\end{subequations}
as $\xi\to\infinity$, where we see the four degrees of freedom, with the asymptotic behaviour imposing three conditions on \refe{eq:inadode}. For the contact line behaviour, in a similar manner we consider
\begin{equation}
 \Psi_1 = \hat{P}_0\xi + \td{\delta}\td{\Psi}_1(\xi), \quad \hat{A}_1 = \hat{S}_0 + \td{\delta}\td{A}_1(\xi),
\end{equation}
and determine the seven eigenmodes as
\begin{align}
 \left.\begin{array}{l}\td{\Psi}_1 = 1\\ \td{A}_1 = 0\end{array}\hspace{-4pt}\right\},
 \left.\begin{array}{l}\td{\Psi}_1 = \xi\\ \td{A}_1 = 0\end{array}\hspace{-4pt}\right\} ,
 \left.\begin{array}{l}\td{\Psi}_1 = \xi^2\\ \td{A}_1 = 0\end{array}\hspace{-4pt}\right\} ,
 \left.\begin{array}{l}\td{\Psi}_1 = \xi^2\ln\xi\\ \td{A}_1 = 0\end{array}\hspace{-4pt}\right\},
 \left.\begin{array}{l}\td{\Psi}_1 = \xi^3\\ \td{A}_1 = 0\end{array}\hspace{-4pt}\right\},
 \left.\begin{array}{l}\td{\Psi}_1 = 0\\ \td{A}_1 = 1\end{array}\hspace{-4pt}\right\},
 \left.\begin{array}{l}\td{\Psi}_1 = 0\\ \td{A}_1 = \xi\end{array}\hspace{-4pt}\right\} .\label{eq:chi0modescl}
\end{align}
The first mode is inconsistent with the asymptotic behaviour and imposes one condition, the other six modes are all consistent and all have free parameters associated with them. With the three conditions from the large-$\xi$ asymptotic behaviour, we now have four conditions, and require three more. Three of the six free parameters at the contact line must thus be fixed by the remaining boundary conditions. The correct specification of the problem requires further terms in the expansion as $\xi\to0$. Proceeding systematically, we find
\begin{subequations}\label{eq:cla1sim}
\begin{align}
\Psi_1 &\sim \hat{P}_0\xi + \hat{P}_1\xi^2\ln\xi + \hat{P}_2\xi^2 + \hat{P}_3\xi^3 + \hat{P}_4\xi^4,\\
\hat{A}_1 &\sim \hat{S}_0 + \hat{S}_1\xi + \left[\frac{\hat{S}_0-\dot{a}}{2\breve{\rho}\bar\beta\bar\tau{\bar\alpha}}
-\frac{3(1+4{\bar\alpha})(2\hat{P}_3 + \hat{S}_1)}{8{\bar\alpha}}
\right]\xi^2,
\end{align}
\end{subequations}
where
\begin{equation}
(1+4{\bar\alpha})\hat{P}_4 =
\frac{\hat{P}_1}{6\bar\beta\bar\tau}+
\frac{16{\bar\alpha}^2-24{\bar\alpha}+1}{32{\bar\alpha}}\hat{P}_3
+\frac{(1+4{\bar\alpha})\dot{a}}{48\breve{\rho}\bar\beta\bar\tau{\bar\alpha}}
-\frac{1+4{\bar\alpha}(1-\breve{\rho})}{48\breve{\rho}\bar\beta\bar\tau{\bar\alpha}}\hat{S}_0
+\frac{(1-4{\bar\alpha})^2}{64{\bar\alpha}}\hat{S}_1
,
\end{equation}
is used for simplicity and we see the six free parameters $(\hat{P}_0,\hat{P}_1,\hat{P}_2,\hat{P}_3,\hat{S}_0,\hat{S}_1)$. 

\paragraph*{\new{Contact line boundary conditions:}} \new{We are now in a position to discuss the boundary conditions at the contact line for mathematical consistency. Firstly, we consider the two agreed contact line conditions \refe{eq:inneradotcl1} and \refe{eq:inneradotcl2}, arising from the mass balance through the contact line and the Young equation. Condition \refe{eq:inneradotcl1} gives the shear stress at the contact line
\begin{equation}
\hat{S}_0 = \frac{2\lrsq{\dot{a}(1-4{\bar\alpha})\breve{\rho}-(1+4\bar\alpha)\hat{P}_1 }}{1+4{\bar\alpha}-8\breve{\rho}{\bar\alpha}},
\end{equation} 
and condition \refe{eq:inneradotcl2} determines
\begin{equation}
6(1+4{\bar\alpha})\hat{P}_3 =
3[\breve{\rho}(1+4{\bar\alpha})-2]\hat{S}_0
- (1+4{\bar\alpha}-8\breve{\rho}{\bar\alpha})\hat{S}_1
- \frac{4}{\bar\beta\bar\tau}\hat{P}_0 + 2[\breve\rho(1+4\bar\alpha)-4]\hat{P}_1.
\end{equation}
There remains one further free parameter to fix for a correctly determined system. Considering the pressure, which corresponds to the second derivative of the droplet height (as seen earlier in \refe{eq:ndnormst}), we have
\begin{equation}
 \ppfrac{\Psi}{\xi} \sim 2\hat{P}_1\ln\xi + 3\hat{P}_1 + 2\hat{P}_2, \quad \mbox{as } \xi\to 0,
\end{equation}
}\new{so that if finite pressure is imposed at the contact line, then this would force $\hat{P}_1=0$, which fixes a free parameter. This behaviour then determines $(\xi\partial^3_\xi \Psi_1)(0)=0$ as suggested earlier.}

\new{As an alternative if finite pressure is not imposed, then the additional condition of Bedeaux\cite{BedeauxExtraCond} and Billingham,\cite{Bill06} \refe{eq:inneradotcl4} requires \begin{equation}
\hat{P}_0  = \frac{[{4{\bar\alpha} \hat{S}_0} + (1-4{\bar\alpha})\dot{a}]\breve{\rho}\lambda}{2\Ca_0},
\end{equation}
fixing $\hat{P}_0$ instead of $\hat{P}_1$. Provided one of these two options is chosen, we will have a well-posed problem, whereas fixing both would then suggest that we would have an over-determined system. It is important to make clear that the logarithmic pressure singularity associated with $\hat{P}_1\neq0$, both for the interface formation model and in an equivalent contact line expansion for the Navier-slip model (discussed later in this section) is not a fatal flaw as it is an integrable singularity, and thus leads to a finite force at the contact line (rather than the infinite force associated with applying the classical no-slip condition). Whether the thermodynamic consideration of the contact line leading to the additional condition may be discounted in preference of finite pressure remains a debate, and was discussed in Sec.~\ref{sec:introduction}.}

\new{We conclude that the equations of the interface formation model are well-posed for the problem considered here, by taking one of the two above options. We will briefly analyse the alternative of using the additional condition, but consider our droplet spreading problem predominantly with the requirement of non-singular pressure at the contact line, which furnishes the problem with a (physical) condition to make it fully determined. This is chosen as a singularity in the pressure at the contact would arise due to the model, rather than from physical origins, and by allowing the (admittedly weak, logarithmic) pressure singularity to exist would suggest that the model is invalid in the immediate vicinity of the contact line. This does however rely on the viewpoint that the additional condition may be dropped or discounted (see the discussion in Sec.~\ref{sec:introduction}).}

\begin{figure}[ht]
\centering
\includegraphics{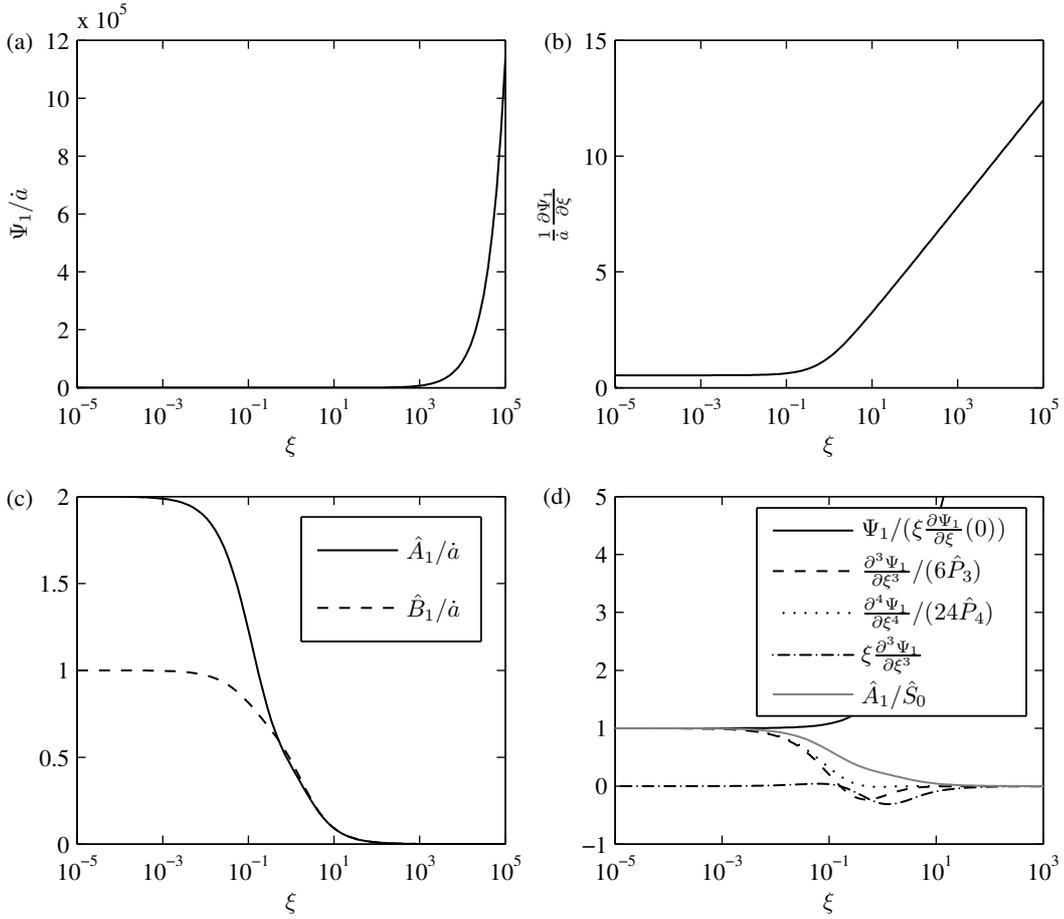}
\caption{Numerical solution of the boundary layer at $O(\dot{a})$ for typical parameter values ${\bar\alpha}=1/12$, $\bar\beta\bar\tau=\breve{\rho}=1$, \new{using the finite contact line pressure condition}. Shown are the $O(\dot{a})$ corrections to (a) the droplet height, (b) the free surface slope, (c) $\hat{A}_1/\dot{a}$, and (d) the solution behaviour as $\xi\to0$, including confirmation of the absence of logarithmic pressure singularity, since $\xi\partial^3_\xi{\Psi_1}\to0$ as $\xi\to0$.}
\label{fig:chi0odeprofile1}
\end{figure}

\begin{figure}[ht]
\centering
\includegraphics{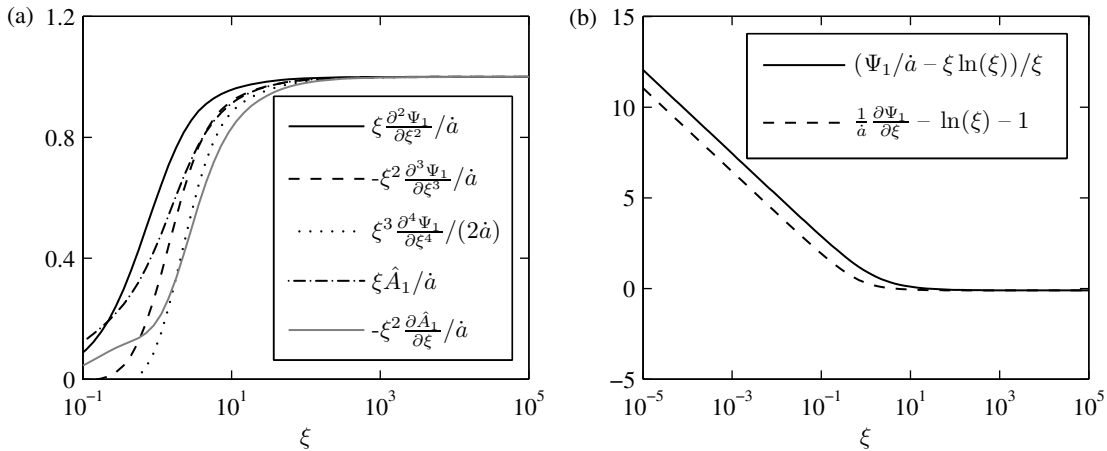}
\caption{
The behaviour of the solution as $\xi\to\infinity$ for typical parameter values ${\bar\alpha}=1/12$, $\bar\beta\bar\tau=\breve{\rho}=1$. (a) Convergence of the second, third and fourth derivatives of the droplet height and of $\hat{A}_1$ and its first derivative, to their expected large $\xi$ behaviour. (b) Determination of the constant $c_0$ from the $\Psi_1$ and $\partial_\xi{\Psi_1}$ behaviours, \new{both using the finite contact line pressure condition}.}
\label{fig:chi0odeprofile2}
\end{figure}

\begin{figure}[ht]
\centering
\includegraphics{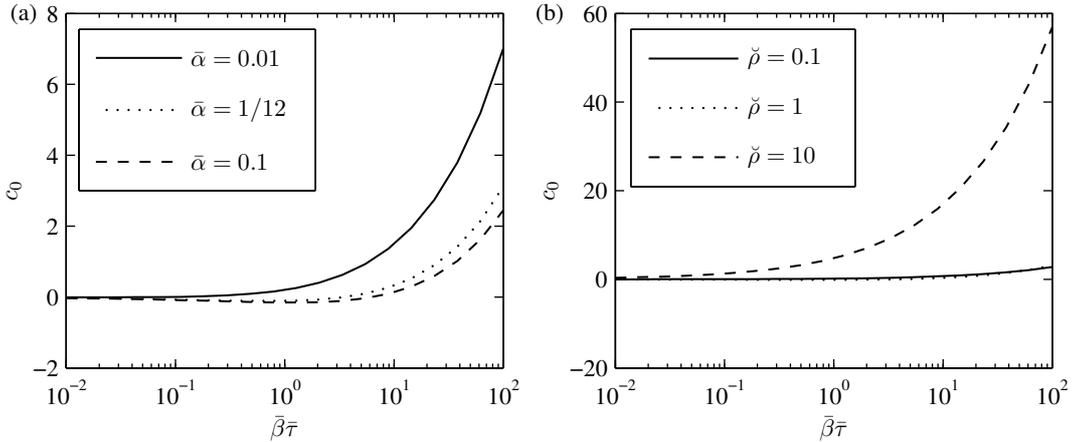}
\caption{The effect of varying $\bar\beta\bar\tau$ on $c_0$ for (a) varying ${\bar\alpha}$ with $\breve{\rho}=1$ fixed, and (b) varying $\breve{\rho}$ with ${\bar\alpha}=1/12$ fixed, \new{both using the finite contact line pressure condition}.}
\label{fig:chi0varyc0}
\end{figure}

\begin{figure}[ht]
\centering
\includegraphics{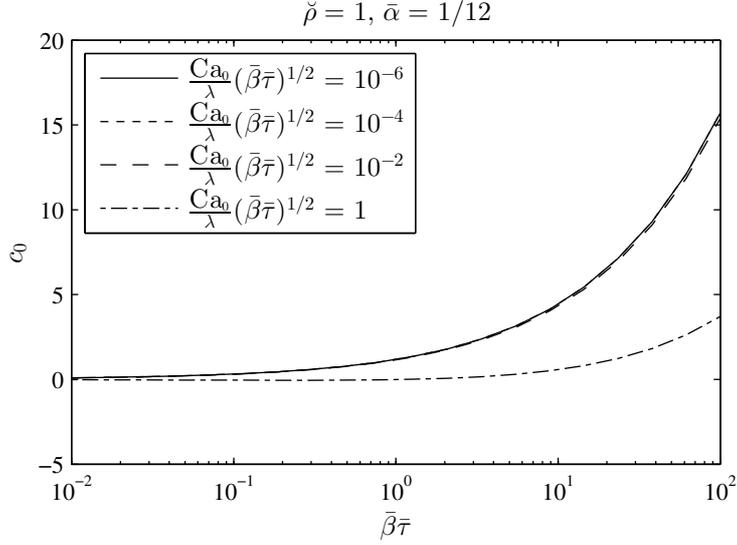}
\caption{\new{The effect of varying $\bar\beta\bar\tau$ on $c_0$ when using the additional condition of Bedeaux and Billingham, and allowing logarithmically diverging pressure at the contact line, for varying ${\Ca_0}{\lambda}^{-1}(\bar\beta\bar\tau)^{1/2}$ with $\breve{\rho}=1$ and ${\bar\alpha}=1/12$ fixed.}}
\label{fig:chi0BedeauvaryCa0}
\end{figure}

\paragraph*{Numerical implementation:}
We consider \refe{eq:inadode} for the variables $\Psi_1/\dot{a},\hat{A}_1/\dot{a},\hat{B}_1/\dot{a}$ so that $\dot{a}$ is removed from the problem, and with fixed values of the parameters \new{${\bar\alpha}$, $\bar\beta\bar\tau$, and $\breve{\rho}$ (and also with $\Ca_0/\lambda$ if using the additional condition)}. This is solved as a two-point boundary value problem (BVP) over the truncated interval $[\xi_0, \xi_\infty]$, with $\xi_0$ and $\xi_\infty$ small and large enough for convergence respectively. Imposing the contact line expansions \refe{eq:cla1sim} supplies four conditions, as determined by our eigenmode analysis, with the matching behaviour \refe{eq:psi1simxiinf} supplying the three remaining conditions and $c_0$ obtained through
\begin{equation}
 c_0 \sim \frac{1}{\xi}\left(\frac{\Psi_1}{\dot{a}} - \xi\ln\xi\right),\quad \mbox{as }\xi\to\infinity.
\end{equation}
These behaviours will be imposed at the interval end points. The BVP is then solved using the MATLAB bvp4c solver which employs a fourth order collocation method based on a three-stage Lobatto IIIa implicit Runge-Kutta formula.\cite{bvp4cref} Figures \ref{fig:chi0odeprofile1}--\ref{fig:chi0odeprofile2} depict typical inner solution behaviours for \new{the finite pressure condition, and }parameter values ${\bar\alpha}=1/12$, $\bar\beta\bar\tau=\breve{\rho}=1$. Figures \ref{fig:chi0odeprofile1}(a)--(c) show the $O(\dot{a})$ corrections to the droplet height, free surface slope, $\hat{A}$ and $\hat{B}$. Figure \ref{fig:chi0odeprofile1}(d) shows the behaviour of the solution as $\xi\to0$, including confirmation that there is no logarithmic pressure singularity through showing $\xi\partial^3_\xi{\Psi_1}\to0$ as $\xi\to0$. Figure \ref{fig:chi0odeprofile2}(a) shows the behaviour of the solution as $\xi\to\infinity$, with convergence to the expected large $\xi$ behaviour, with Fig. \ref{fig:chi0odeprofile2}(b) showing the determination of $c_0$ from the $\Psi_1$ and $\partial_\xi{\Psi_1}$ behaviours. The effect on $c_0$ when varying $\bar\beta\bar\tau$ for selected ${\bar\alpha}$ and $\breve{\rho}$ is given in Fig. \ref{fig:chi0varyc0}. \new{To also demonstrate the effect of using the additional condition on the spreading rate through the value of $c_0$, we include Fig. \ref{fig:chi0BedeauvaryCa0}, which has varying $\bar\beta\bar\tau$ for fixed ${\bar\alpha}$ and $\breve{\rho}$, and selected ${\Ca_0}{\lambda}^{-1}(\bar\beta\bar\tau)^{1/2}$. We choose to vary this combination as $\Ca_0$ and $\lambda$ appear only as ${\Ca_0}{\lambda}^{-1}$ in the governing equations, and combining them in this way with $\bar\beta\bar\tau$ removes the dependence on the long-wave parameter $\epsilon$. Based on the nondimensional numbers in \refe{eq:paramsizes}, we have ${\Ca_0}{\lambda}^{-1}(\bar\beta\bar\tau)^{1/2}\approx 2^{-1/2}\times10^{-4}$.}

\begin{figure}[ht]
\centering
\includegraphics{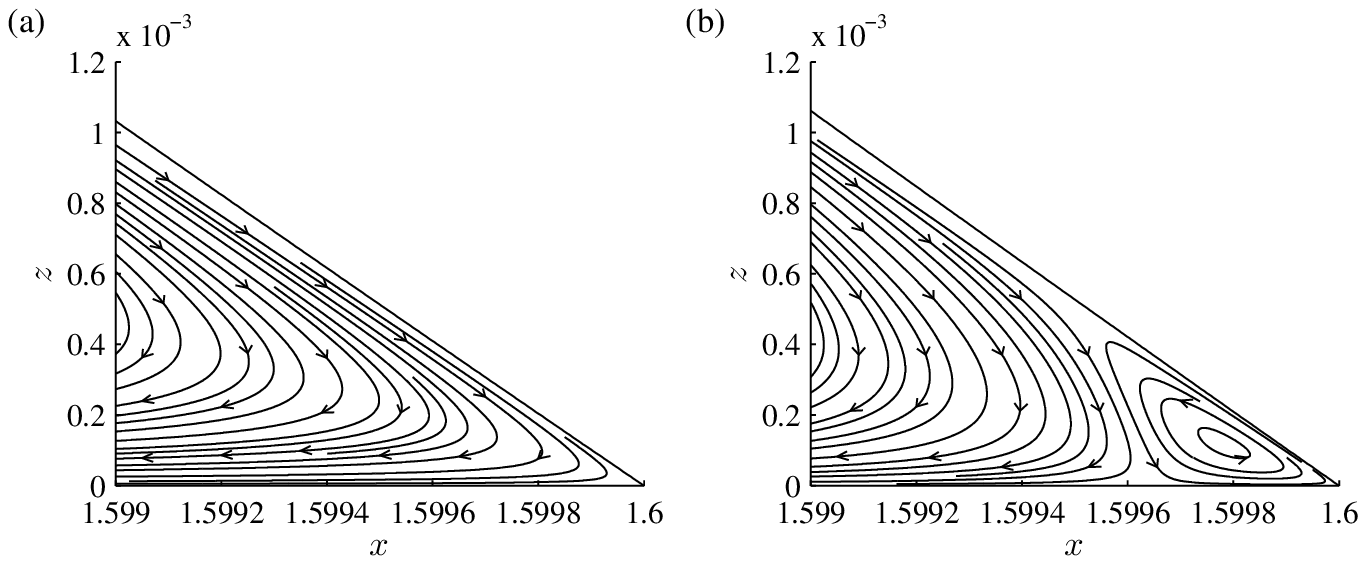}
\caption{Streamlines for two flow situations arising from different sets of parameter values, in the frame of reference of the moving contact line. (a) shows the expected flow situation with no region containing closed streamlines using parameter values $\bar\beta=10^3$, $\bar\alpha=1/12$, $\bar\beta\bar\tau=\breve{\rho}=1$. (b) has parameter values chosen as $\bar\beta=10^3$, $\bar\alpha=1/12$, $\breve{\rho}=10$, $\bar\beta\bar\tau=0.1$, and shows a situation where the flow-induced Marangoni effect occurs. \new{Both flow visualisations use the finite contact line pressure condition.}}
\label{fig:maran}
\end{figure}

Whilst not undertaking an exhaustive analysis of the flow fields in the vicinity of the contact line, we do include two situations for comparison, in Fig.~\ref{fig:maran}. Streamlines are shown in the frame of reference of the moving contact line, and plotted in outer variables. In (a) we show a typical flow situation with parameter values $\bar\beta=10^3$, $\bar\alpha=1/12$, $\breve{\rho}=\bar\beta\bar\tau=1$ where the behaviour is as found for slip models, with the fluid flowing down the free surface towards the contact line and away along the solid surface. In (b) parameter values are chosen to be $\bar\beta=10^3$, $\bar\alpha=1/12$, $\breve{\rho}=10$, $\bar\beta\bar\tau=0.1$. This is a situation where the flow-induced ``Marangoni effect" occurs (see \S 5.2.3 of the monograph of Shikhmurzaev for details of this effect\cite{ShikhBook}), being caused by the surface-tension gradient driven by the flow field in the region of the contact line. \new{Both flow visualisations use the finite contact line pressure condition.}

\begin{figure}[ht]
    \centering
    \includegraphics{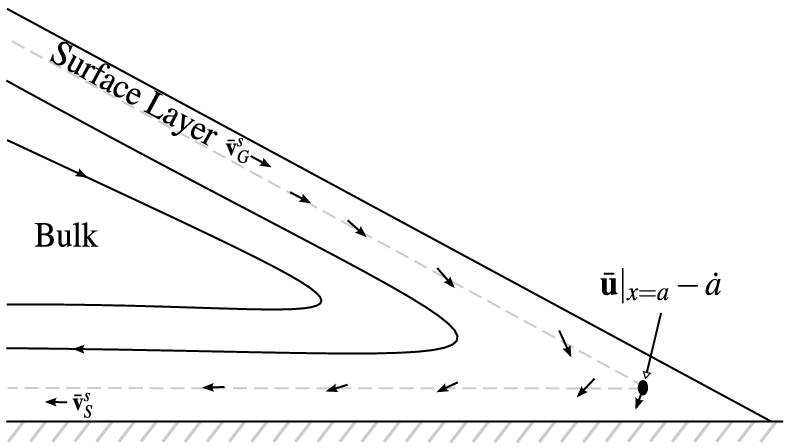}
    \caption{Diagram of rolling motion in the frame of reference of the contact line, with interfaces shown schematically as layers of finite thickness. The contact line created by the bulk is shown with a circle, where the velocity is shown as non-zero. This occurs for the interface formation model with $\chi_m=1$, but not when $\chi_m=0$ or with slip models, where a stagnation point occurs.} \label{fig:rolling}
\end{figure}

\paragraph*{A comparison to slip with a velocity dependent contact angle:}
Matching of the interface formation model equations between inner and outer regions has given rise to a modified equation for the spreading rate when compared to a Navier-slip model with $\theta_d=\theta_s$. Given that the interface formation model allows the \new{microscopic} dynamic contact angle to vary with contact line velocity, it is of interest to see how a similar behaviour would change the Navier-slip model. As an example, we consider the contact line condition
\begin{equation}
 \left(\pfrac{\Bah}{\Bax}(a)\right)^2 = 1 + 2\dot{a} c_{NS}. \label{eq:varCAvel}
\end{equation}
The equations of the Navier-slip model in the inner region give $\Psi_{0NS}=\xi$, and at $O(\dot{a})$ we have
\begin{equation}
 \Psi_{1NS} = \frac{\dot{a}}{2}\left[ (1+\xi)^2\ln(1+\xi)-\xi(1-2c_{NS}+\xi\ln\xi) \right], \quad \mbox{having applied}\quad
 \pfrac{\Psi_{1NS}}{\xi}(0) = c_{NS}\dot{a},
\end{equation}
and where other arbitrary constants have been found using $\Psi_{1NS}/\xi^2\to0$ as $\xi\to\infinity$ and $\Psi_{1NS}(0)=0$. Thus the behaviour as $\xi\to\infinity$ is found to be
\begin{equation}
 \pfrac{\Psi_{1NS}}{\xi} \sim \dot{a}\ln\left( \mye^{1+c_{NS}}\xi \right), \qquad \mbox{leading to}\qquad
3\dot{a}a^6\ln\left( \frac{\beta_{NS}\;\mye^{2-c_{NS}}}{2a} \right) = a^6-27,\label{eq:NSspr}
\end{equation}
an equation for the spreading rate. This shows that the variation of $\theta_d$ with contact line velocity impacts the spreading rate in the same way as found for the interface formation model, in equation \refe{eq:adotina}. In the Navier-slip case, we can show analytically that the spread rate is affected precisely by the velocity dependence of the \new{microscopic} contact angle, whereas the constant $c_0$ must be found numerically in the interface formation model. We note that the Navier-slip model is not the only slip model that can be reduced to this spreading rate equation. The slip model of Ruckenstein and Dunn\cite{RuckDunn} yields $ \Bau = (\beta_{RD}^2/{\Bah})\partial_{\Baz} {\Bau}$, on $\Baz=0$ in the long-wave approximation.\cite{HaleyMiksis,SavvaPrecursorSlip} This has the inner solution
\begin{equation}
 \Psi_{1RD} = \frac{\dot{a}}{2}\left[ \xi\ln(1+\xi^2)
+(1-\xi^2)\arctan\xi
+\frac{\xi}{2}(\pi\xi-2+4c_{RD}) \right],
\end{equation}
where we have imposed $\partial_\xi \Psi_{1RD}(0) = c_{RD}\dot{a}$, and we find that \refe{eq:NSspr} holds with $c_{NS}=c_{RD}$. A comparison between slip and precursor film models is given by Savva and Kalliadasis,\cite{SavvaPrecursorSlip} detailing a wider class of models which also reduce to equivalent spreading dynamics. We note that the Navier-slip model still has logarithmically singular pressure at the contact line, whereas the Ruckenstein and Dunn model has finite pressure since
\begin{equation}
 \ppfrac{\Psi_{NS}}{\xi} \sim -\dot{a}\ln\xi + \dot{a}\xi + O(\xi^2),
 \qquad
 \ppfrac{\Psi_{RD}}{\xi} \sim \frac{\dot{a}\pi}{2}-\dot{a}\xi
 + O(\xi^2), \qquad \mbox{as }\xi\to0.
\end{equation}
The interface formation model \new{is also able to alleviate} the pressure singularity as discussed earlier.
Another feature of interest is the contact line velocity. For both slip models and for the interface formation model with $\chi_m=0$ we have $\Bau|_{\Bax=a}=\dot{a}$. This suggests that when considering a frame of reference moving with the contact line, the contact line is a stagnation point, and rolling motion does not occur. We next consider the inclusion of the extra terms when $\chi_m=1$, and will find that this stagnation point is removed. This then allows for the rolling motion seen experimentally as discussed in the introduction, with an illustrative schematic in Fig.~\ref{fig:rolling}.

%
%

\section{Inclusion of additional mass transfer terms, $\chi_m=1$}
\label{sec:chim1}

Returning to the nondimensional equations under the long-wave assumption \refe{eq:ndbulk}--\refe{eq:ndcllast2}, we now consider the case where $\chi_m=1$ for the modern version of the interface formation model of Shikhmurzaev.\cite{ShikhBook} The parameter $\bar{R}_0 \approx 1.5\times10^{-4}\epsilon^{-2}$, based on the sizes of the nondimensional numbers given earlier in \refe{eq:paramsizes} for water at room temperature, so that $\bar R_0 = O(1)$ if $\epsilon=O(1.2\times 10^{-2})$. Analysis considering $\bar R_0 = O(1)$ should thus also be performed.

The differences to the previous analysis for $\chi_m=0$ appear only in the equation for $\Bav$ on the solid surface and in the kinematic condition on the gas surface. In a similar manner to the $\chi_m=0$ case, we thus determine
\begin{equation}
 \Bau =-\frac{3\Baz^2}{2}{\pppfrac{\Bah}{\Bax}}  +  3A\Baz +  B,\qquad
 \Bav = \frac{\Baz^3}{2}{\ppppfrac{\Bah}{\Bax}}  - \frac{3\Baz^2}{2}\pfrac{A}{\Bax} - \Baz \pfrac{B}{\Bax} + \frac{\bar{R}_0}{3}\BrS,
\end{equation}
and these additions manifest themselves in the free surface equation
only, which now becomes
\begin{equation}
 \pfrac{\Bah}{\Bat} + \pfrac{}{\Bax}\left(-\frac{1}{2}\Bah^3 \pppfrac{\Bah}{\Bax}
+ \frac{3}{2}A\Bah^2 + B\Bah\right)
= \bar{R}_0 \mathcal{D}  ,\label{eq:chi1eveq}
\end{equation}
where $\mathcal{D}$ satisfies \refe{eq:chi0rhoSG}. $A(\Bat,\Bax)$ and $B(\Bat,\Bax)$ once again satisfy \refe{eq:chi0AB1} and \refe{eq:chi0AB2} and also at the contact line the conditions \refe{eq:chi0ocl3} are again applicable, but with the modified contact line velocity $ \bar{U}_{CL} = B - \bar R_0\left(\partial_{\Bax}\Bah\right)^{-1}\mathcal{D} $. Mass conservation is determined by \refe{eq:massconR0}, suggesting that mass transfer to the surface layers drives the actual contact line velocity from that created by the bulk. Given that in the outer region of the droplet $\bar\tau\ll1\ll\bar\beta$, then all additional terms are negligible and the outer region solution for $\chi_m=1$ follows as in the $\chi_m=0$ case.

\paragraph*{The inner region:}
We next consider the inner region with the scalings \refe{eq:innerscals} and the quasistatic expansions \refe{eq:inquas}. For the leading order solution we find $\Psi_0 = \xi$, following a similar argument as in the $\chi_m=0$ case. Considering the next order in the governing equations, we have
\begin{equation}\label{eq:chi1adot1}
\dot{a}_0 - \pfrac{}{\xi}\left(\frac{1}{2}\xi^3 \pppfrac{\Psi_1}{\xi}
+ \frac{3}{2}\hat{A}_1\xi^2 + {\hat{B}_1}\xi\right) = \bar{R}_0 \mathcal{D},
\end{equation}
with \refe{eq:inneradot2}--\refe{eq:inneradot3}, and where $\mathcal{D}$ is given by \refe{eq:chi0inOadotrhoGS}.
At $\xi=0$ the contact line conditions are
\begin{subequations}\label{eq:chi1:inneradotlast}
\begin{align}
 \hat{B}_1 + 4\bar\alpha(\hat A_1-\hat B_1)
 &= \frac{1+4{\bar\alpha}}{2\breve{\rho}}\left[ \xi\pppfrac{\Psi_1}{\xi}+\hat A_1\right]
 + \frac{2\left( 1+ \breve{\rho}\right)\bar R_0}{\breve{\rho}}\pfrac{\Psi_1}{\xi} ,
\\
  -\mathcal{D} &= \pfrac{\Psi_1}{\xi},
\\ \label{eq:chi1:inneradotlastex}
  4{\bar\alpha}(\hat B_1-\hat A_1) - \hat B_1 &= 2\mathcal{D}\left[\bar{R}_0 + \frac{\Ca_0}{\lambda\breve{\rho}} \right],
\end{align}
\end{subequations}
with $\Psi_1(0)=0$. Once again we require $\Psi_1/\xi^2\to0$, $\hat{A}_1\to0$, $\hat{B}_1\to0$ as $\xi\to\infinity$ to match to the outer solution. Studying the extra terms in this $\chi_m=1$ situation, it may be seen that they will not contribute at leading order as $\xi\to\infinity$, and an asymptotic expansion confirms that the matching results of \refe{eq:A1B1xi2inf} still hold, with differences occurring only at higher order. Obtaining further terms for the matching of $\hat{A}$ and $\hat{B}$ again follows as for $\chi_m=0$ so that the matching behaviours are in agreement with \refe{eq:matchfrominner}.
Given the outer region is in agreement for $\chi_m=0$ and $\chi_m=1$, as is the leading order behaviour for $\xi\to\infinity$ from the inner region, then the matching and the result for the rate of droplet spreading \refe{eq:adotina} will be the same.

Returning to the $O(\dot{a}_0)$ inner region equations \refe{eq:chi1adot1}--\refe{eq:chi1:inneradotlast} with \refe{eq:inneradot2}--\refe{eq:inneradot3} and \refe{eq:chi0inOadotrhoGS}, we now wish to find a full solution to complete our asymptotic description and obtain an $O(\dot{a}_0)$ correction to the \new{microscopic} contact angle. We drop the subscript $0$ from $a_0$ and $\dot{a}_0$ for the remainder of this section as we have found that corrections to the radius do not enter into our analysis, as for $\chi_m=0$. Equation \refe{eq:chi1adot1} may be solved to give
\begin{equation}
 \hat{B}_1 = \frac{2(\dot{a}\xi+\hat{K}_\chi) - \xi^2(3\hat{A}_1+\xi\partial^3_\xi{\Psi_1})}{2\xi+\bar\beta\bar\tau\bar{R}_0(\breve{\rho}(1+4{\bar\alpha})+2)}
-\bar\beta\bar\tau\bar{R}_0
\frac{ (12\xi+1+4{\bar\alpha}-8\breve{\rho}{\bar\alpha}) \hat{A}_1 + (6\xi+1+4{\bar\alpha}) \xi\partial^3_\xi{\Psi_1}}{4\xi+2\bar\beta\bar\tau\bar{R}_0(\breve{\rho}(1+4{\bar\alpha})+2)}
,\label{eq:b1solchi1}
\end{equation}
where $\hat{K}_\chi$ is a constant of integration which we cannot remove in this case, as we have
\begin{equation}
 \hat{B}_1(0) = \frac{4\hat{K}_\chi -
 \bar\beta\bar\tau\bar{R}_0(1+4{\bar\alpha}-8\breve{\rho}{\bar\alpha})\hat{A}_1(0)}{\bar\beta\bar\tau\bar{R}_0(4+2\breve{\rho}(1+4{\bar\alpha}))},
\end{equation}
which is finite (assuming $\hat{A}_1(0)$ is finite) as required and as such does not remove the arbitrary constant. After substantial simplification, the first and second contact line conditions together reduce to
\begin{equation}
 \hat{K}_\chi = \bar\beta\bar\tau\bar{R}_0\dot{a}(1+\breve{\rho}), \qquad \mbox{and} \qquad
 \hat{A}_1(0) = \frac{2\dot{a}(1-4{\bar\alpha})\breve{\rho}}{1+4{\bar\alpha}-8\breve{\rho}{\bar\alpha}} -2\bar{R}_0
 \frac{2+\breve{\rho}(1+4{\bar\alpha})}{1+4{\bar\alpha}-8\breve{\rho}{\bar\alpha}}\pfrac{\Psi_1}{\xi}(0),
\end{equation}
so that $\hat{B}_1(0) = \dot{a} + \bar{R}_0\partial_\xi{\Psi_1}(0)$. This also confirms $\bar{U}_{CL} = \hat{B}(0)-\bar{R}_0\partial_\xi{\Psi_1}(0) = \dot{a}$, differing from the bulk contact line velocity $\hat{B}_1(0)$, and allowing rolling motion to take place. We find through a similar eigenmode analysis as for $\chi_m=0$ that the third contact line condition \refe{eq:chi1:inneradotlastex} is not required, \new{provided a finite pressure at the contact line is instead imposed}. We are left with a system of two ODEs of seventh order, which is obtained from \refe{eq:inneradot2}--\refe{eq:inneradot3}, \refe{eq:chi0inOadotrhoGS}, and \refe{eq:chi1:inneradotlast} and substituting in the above results. We will not record this here due to its cumbersome size.

\begin{figure}[ht]
    \centering
    \includegraphics{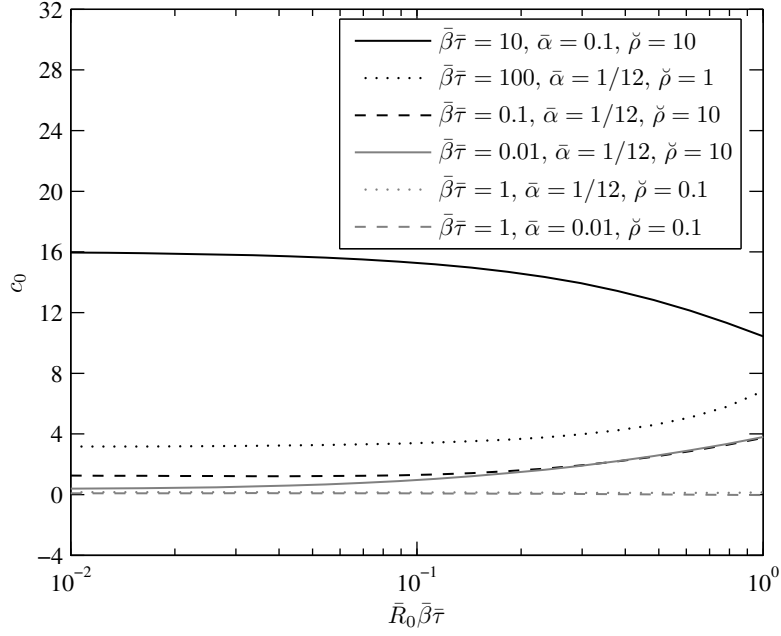}
    \caption{The effect on $c_0$ when varying $\bar{R}_0\bar\beta\bar\tau$ for selected values of the other parameters \new{using the finite contact line pressure condition}.}
\label{fig:varyR0}
\end{figure}

\begin{figure}[ht]
\centering
\includegraphics{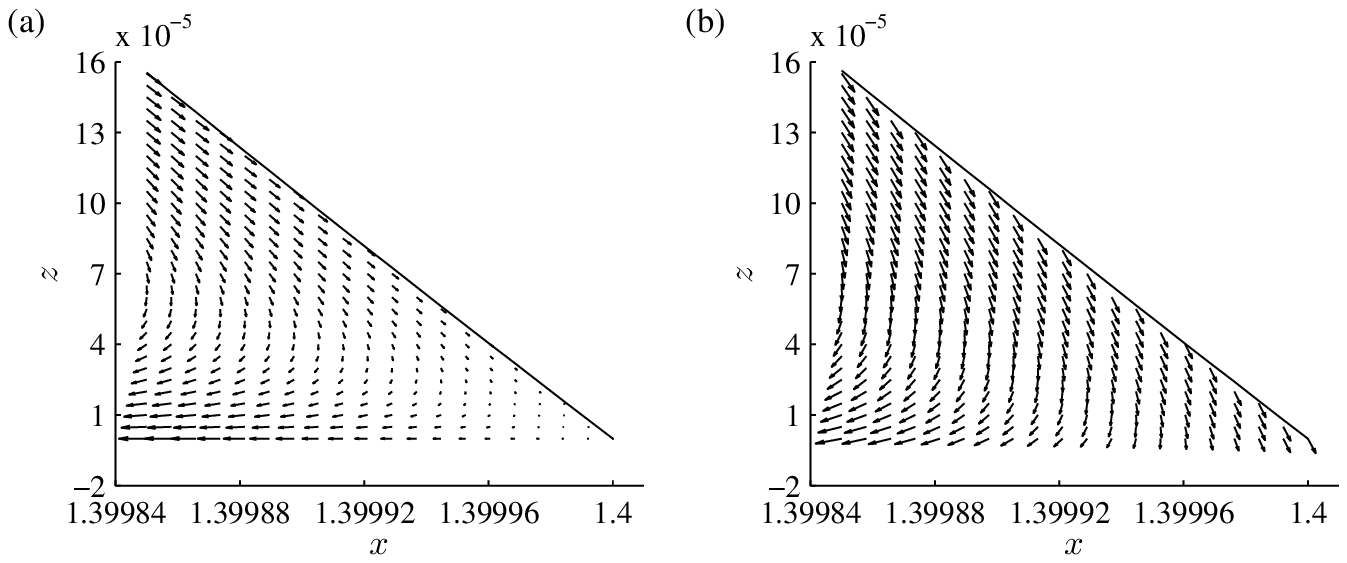}
\caption{Velocity plots in the region close to the contact line of the two formulations of the interface formation model, (a) $\chi_m=0$ and (b) $\chi_m=1$, in the frame of reference of the contact line. Parameter values chosen as $\bar\beta=10^3$, $\bar\alpha=1/12$, $\breve{\rho}=0.1$, $\bar\beta\bar\tau=1$, with additionally $\bar{R}_0=1/3$ in (b). Plots shown in outer variables for perceptibility and to relate to Fig.~\ref{fig:rolling}, \new{and both use the finite contact line pressure condition}.}
\label{fig:rollingvelplots}
\end{figure}

Numerical implementation of this system with the solution for $\hat{B}_1$ from \refe{eq:b1solchi1} follows as for the $\chi_m=0$ situation, and we refer to the details of Sec. \ref{sec:fullOadotchi0}. Figure \ref{fig:varyR0} illustrates the effect on $c_0$ when varying $\bar{R}_0\bar\beta\bar\tau$ for selected values of the parameters ($\bar\beta\bar\tau$,${\bar\alpha}$,$\breve{\rho}$). We choose to vary $\bar{R}_0\bar\beta\bar\tau$ as based on the nondimensional numbers in \refe{eq:paramsizes}, we have $\bar{R}_0\bar\beta\bar\tau\approx 1/3$ with no dependence on the long-wave parameter $\epsilon$. As $\bar{R}_0\bar\beta\bar\tau\to0$, $c_0$ approaches the value from the $\chi_m=0$ case. Figure \ref{fig:rollingvelplots} compares the velocity fields close to the contact line of the two formulations of the interface formation model, (a) $\chi_m=0$ and (b) $\chi_m=1$. The same parameter values of $\bar\beta=10^3$, $\bar\alpha=1/12$, $\breve{\rho}=0.1$, $\bar\beta\bar\tau=1$, are used in both figures with additionally $\bar{R}_0=1/3$ for the $\chi_m=1$ formulation. The predicted low velocity region is seen in (a) for $\chi_m=0$, with the contact line a stagnation point. This prevents \new{nanoscale} rolling motion as a fluid particle on the surface will not reach the contact line in finite time. This issue is resolved in the model with $\chi_m=1$, as seen in (b). \new{We note that mass conservation for bulk and surface layers is guaranteed from the formulation of the model, as the original equations lead to the mass balance \refe{eq:cofmdim} and the first contact line condition in \refe{prenondim:clcond}.}

\section{Numerical analysis}
\label{sec:numerics}

Returning to the full problem in both model formulations, we wish to solve the governing PDEs numerically to verify our asymptotic analysis. We will first consider the $\chi_m=0$ case, with the $\chi_m=1$ case then following a similar procedure \new{and by considering the model with the requirement of finite pressure at the contact line.}

\subsection{Numerical solution of the full problem for $\chi_m=0$}
\label{sec:numchi0}
We now return to the governing equations of the full boundary value problem for $\chi_m=0$ in terms of the non-moving coordinate $y$, where $\Bax = a y$, given in \refe{eq:actualprob}--\refe{eq:outrgrs}, and with conditions:
\begin{equation}
 \Bah(1) = 0, \qquad \partial_y{\Bah}(0)=0,
\qquad
 A(0) = B(0) = 0, \qquad \int_0^1 \Bah dy = {a}^{-1}.\label{eq:numchi0eotherconds}
\end{equation}
Considering the evolution equation in \refe{eq:eq1actualprob} as $y\to 1$, we find $\dot{a} =  B(1)$ as expected, where we have used the requirement that $\left(\Bah\partial^3_y{\Bah}\right)(1)=0$ to alleviate the logarithmic pressure singularity, and that we also do not want a singularity in $A$ at the contact line. This confirms $\bar{U}_{CL} = \dot{a}$. To further simplify the equations for our numerical scheme, we note that we may rewrite $A$ and $B$ to avoid the calculation of the fifth derivative of the free surface. Consequently, we make the substitutions
\begin{equation}\label{eq1:subsrem5th2}
 \breve{A} = A -
 \frac{(6\bar\beta\Bah+1+4\bar\alpha)(1+4\bar\alpha)a^{-3}\Bah\partial^3_y{\Bah}}
{\breve\alpha+12(1+4\bar\alpha)\Bah\bar\beta}, \qquad
 \breve{B} = \bar\beta B -  \frac{4\bar\alpha(6\bar\beta\Bah+1+4\bar\alpha)a^{-3}\Bah\partial^3_y{\Bah}}
{\breve\alpha+12(1+4\bar\alpha)\Bah\bar\beta},
\end{equation}
where $\breve\alpha=1+24\bar\alpha+16\bar\alpha^2$ for simplicity, so that the governing equations become
\begin{subequations}
\begin{align}
\pfrac{\Bah}{\Bat} &= \frac{\dot{a}y}{a}\pfrac{\Bah}{y} - \pfrac{}{y}\left[
\frac{3\breve{A}\Bah^2}{2a} + \frac{\breve B\Bah}{\bar\beta a}
+
\frac{(3\bar\beta^2\Bah^2 + 4\bar\alpha)(1+4\bar\alpha)\Bah^2
+ \breve\alpha\bar\beta\Bah^3  }
{[12(1+4\bar\alpha)\bar\beta^2\Bah+\breve\alpha\bar\beta]a^4} \pppfrac{\Bah}{y}
\right] ,
\label{eq:num1chi0e1}
\\\label{eq:num1chi0e2}
\breve A &= \frac{\bar\tau\breve\rho\bar\alpha}{\bar\beta a^2} \ppfrac{\breve A}{y} + \breve B -\frac{\bar\tau\breve\rho(1+4\bar\alpha)}{4\bar\beta a^2} \ppfrac{\breve B}{y} - \frac{6\bar\beta\Bah+1+4\bar\alpha}{\breve\alpha+12(1+4\bar\alpha)\Bah\bar\beta}\frac{\Bah}{a^3}\pppfrac{\Bah}{y},
\\
\breve A &=
\frac{\bar\tau}{\bar\beta a^2}\left[
{6}\bar\beta\pfrac{\Bah}{y}\pfrac{\breve A}{y} + \frac{12\Bah\bar\beta+1+4\bar\alpha}{4}\ppfrac{\breve A}{y}
+\ppfrac{\breve B}{y}
 +3\bar\beta{\breve A}\ppfrac{\Bah}{y} \right]
\nonumber\\
&\qquad\qquad\qquad\qquad\qquad\qquad\qquad\qquad\qquad\qquad\qquad+ \frac{2}{a^3}\frac{3(1+4\bar\alpha)\bar\beta\Bah+8\bar\alpha}{\breve\alpha+12(1+4\bar\alpha)\Bah\bar\beta}
\Bah\pppfrac{\Bah}{y}
,\label{eq:num1chi0e3}
\end{align}
\end{subequations}
with at $y=1$
\begin{gather}
\breve B =
\frac{1+4\bar\alpha-8\breve\rho\bar\alpha}{2\breve\rho(1-4\bar\alpha)}\breve A ,
\nonumber\\
\left(\pfrac{\Bah}{y}\right)^2 =
a^2 +
\frac{a\bar\tau}{2\bar\beta} \pfrac{}{y} \left[
\left({1+4\bar\alpha-8\breve\rho\bar\alpha+12\bar\beta\Bah} \right)\breve A +
2(\breve\rho (1+4\bar\alpha)+2)\breve B \right] ,\label{eq:fn:bc1}
\end{gather}
and
\begin{gather}
 \left(\Bah\partial^3_y{\Bah}\right)(1)=0, \quad \Bah(1) = 0, \quad \partial_y{\Bah}(0)=0, \quad \breve A(0) = \breve B(0) = 0, \quad
 \int_0^1 \Bah dy = {a}^{-1}, \quad {\breve B(1)} = \dot{a}\bar\beta.\label{eq:fn:otherbcs}
\end{gather}
For a given free surface profile at a particular point in time, we may solve for $\breve A$ and $\breve B$ using the second two governing equations \refe{eq:num1chi0e2}--\refe{eq:num1chi0e3} and boundary \new{conditions \refe{eq:fn:bc1}}, along with $\breve A(0) = \breve B(0) = 0$. We then have the evolution equation \refe{eq:num1chi0e1}, which contains a fourth order spatial derivative of $\Bah$ as in the Navier-slip case, along with the remaining five conditions of \refe{eq:fn:otherbcs}. Comparing the conditions with the Navier-slip case, we have here $\left(\Bah\pppilfrac{\Bah}{y}\right)(1)=0$ and $\dot{a} = {\breve B(1)}/{\bar\beta}$, whereas for Navier-slip we would have $\left(\pilfrac{\Bah}{y}\right)^2(1)=a^2$, and $\dot{a} = \beta_{NS}a^{-3}\left(\Bah\pppilfrac{\Bah}{y}\right)(1)$. This would suggest that the additional condition, not listed above, is not required \new{when enforcing finite pressure at the contact line}---in agreement with our asymptotic analysis.

To guarantee mass conservation we employ the transformation of Savva and Kalliadasis\cite{Savva09} to solve for the integral of the free surface. We consider
\begin{equation}
 G = \int_0^y \Bah(y') dy',\label{eq:GasH}
\end{equation}
so that the PDE \refe{eq:num1chi0e1} becomes
\begin{equation}
\pfrac{G}{\Bat} = \frac{\dot{a}}{a}\left[y\pfrac{G}{y} - G\right]
- \frac{3\breve{A}\Bah^2}{2a} - \frac{\breve B\Bah}{\bar\beta a}
-
\frac{(3\bar\beta^2\Bah^2 + 4\bar\alpha)(1+4\bar\alpha)\Bah^2
+ \breve\alpha\bar\beta\Bah^3  }
{[12(1+4\bar\alpha)\bar\beta^2\Bah+\breve\alpha\bar\beta]a^4}   \pppfrac{\Bah}{y},
\end{equation}
where we have integrated the evolution equation with respect to $y$. Now at $y=1$
\begin{gather}
 G(1) = {1}/{a}, \qquad \partial_y{G}(1) = 0, \qquad \breve B(1)={\bar\beta}\dot{a},
\end{gather}
where the value of $G(1)$ \new{comes from the definition of $G$ in \refe{eq:GasH} and the volume condition in \refe{eq:numchi0eotherconds} and we then automatically satisfy}
\begin{equation}
\int_0^1 \Bah dy = G(1) - G(0) = 1/{a}.
\end{equation}
We should point out a difference in the setup of the problem in comparison to the Navier-slip model. The contact line velocity condition $\dot{a} = {\breve B(1)}/{\bar\beta}$ may still be applied here, whereas the equivalent condition in Navier-slip is of the same order as in the PDE, and can no longer be used. Instead, we lose $\left(\Bah\pppilfrac{\Bah}{y}\right)(1)=0$ in a similar manner, and then have the correct number of boundary conditions. The numerical solution of the evolution equation is based on spectral differentiation in space and adaptive, semi-implicit time stepping, following similar ideas from the scheme outlined in the Appendix of the study by Savva and Kalliadasis.\cite{Savva09} \new{We note that along with the droplet radius behaviour we are also interested here in the variation of the microscopic contact angle, which given its dependence on the contact line velocity, is very sensitive to the initial condition imposed. Here, we specify an arbitrary droplet profile which has agreement with
the leading order outer solution and with boundary layers of width $O(\bar\beta^{-1})$, which is then allowed to briefly relax towards a quasistatic regime to provide the initial condition imposed. This then allows for a fair comparison between asymptotic and numerical results by minimising the impact of the specified initial condition.}

\begin{figure}[ht]
    \centering
    \includegraphics{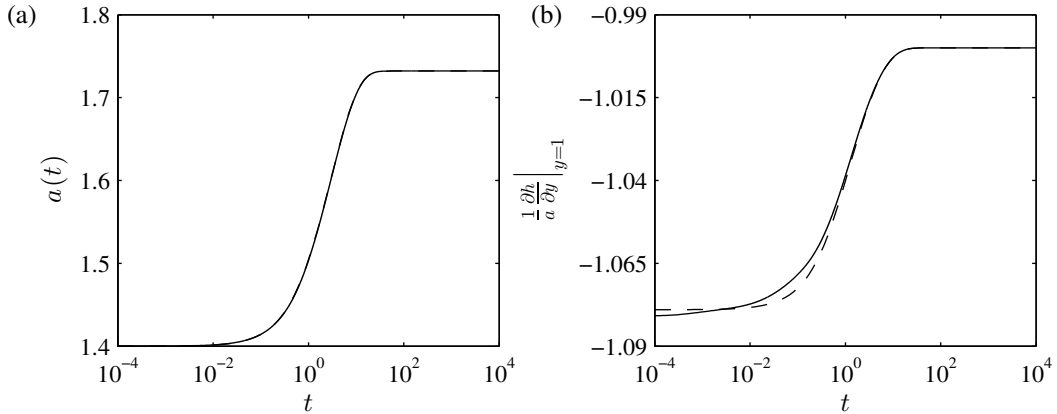}
    \caption{Interface formation model with $\chi_m=0$: \new{Evolution of (a) the droplet fronts, and (b) the microscopic contact angle for initial droplet radii $a(0)=1.4$, with parameter values are chosen to be $\bar\beta=10^3$, $\bar\beta\bar\tau=1$, ${\bar\alpha}=1/12$, $\breve{\rho}=1$. Dashed lines use the asymptotic equations from matching with $c_0 = -0.1$ and $\hat{P}_0/\dot{a} = 0.54$. Solid lines give the solution of the PDEs, and in (a) the lines are nearly indistinguishable.}}
\label{fig:origpde}
\end{figure}

\begin{figure}[ht]
    \centering
    \includegraphics{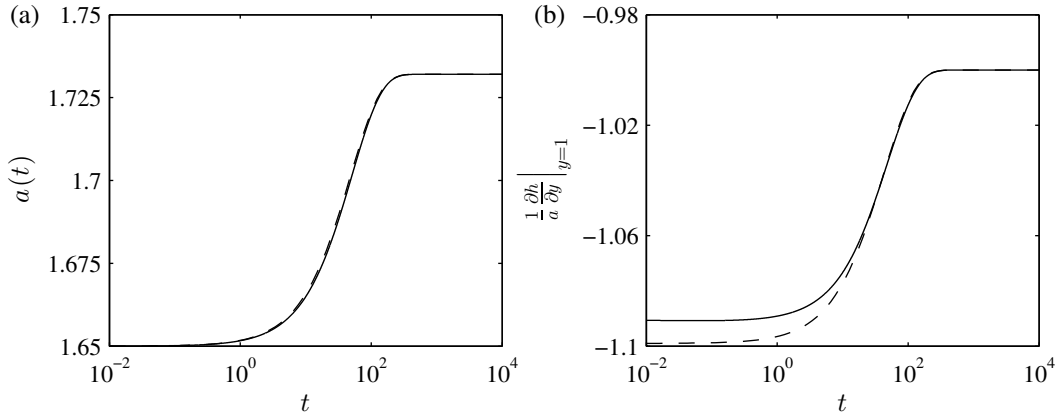}
    \caption{Interface formation model with $\chi_m=0$: \new{Evolution of (a) the droplet fronts, and (b) the microscopic contact angle for initial droplet radius $a(0)=1.65$, with parameter values chosen to be $\bar\beta=10^3$, $\bar\beta\bar\tau=100$, ${\bar\alpha}=1/12$, $\breve{\rho}=10$. Dashed lines use the asymptotic equations from matching with $c_0 = 56.96$ and $\hat{P}_0/\dot{a} = 55.4$. Solid lines give the solution of the PDEs.}}
\label{fig:bigc0chi0a0v16}
\end{figure}

\begin{figure}[ht]
    \centering
    \includegraphics{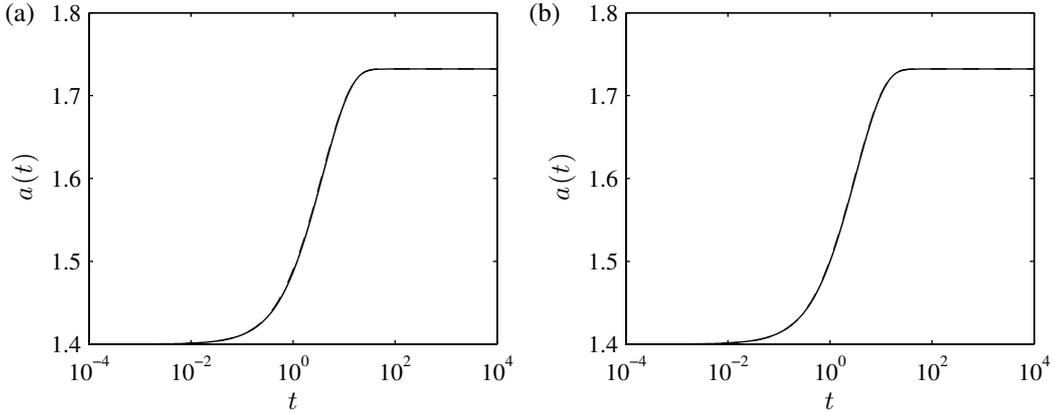}
    \caption{Interface formation model with $\chi_m=0$: \new{Evolution of the droplet fronts for initial droplet radii and parameter values (a) $a(0)=1.4$, $\bar\beta=10^3$, $\bar\beta\bar\tau=0.1$, ${\bar\alpha}=1/12$, $\breve{\rho}=10$ and (b) $a(0)=1.4$, $\bar\beta=10^3$, $\bar\beta\bar\tau=1$, ${\bar\alpha}=1/12$, $\breve{\rho}=0.1$. Dashed lines use the asymptotic equations from matching with (a) $c_0 = 1.28$ and (b) $c_0=0.16$. Solid lines are from the solution of the PDEs and are nearly indistinguishable.}} 
\label{fig:chi0asmaran}
\end{figure}

\begin{figure}[ht]
    \centering
    \includegraphics{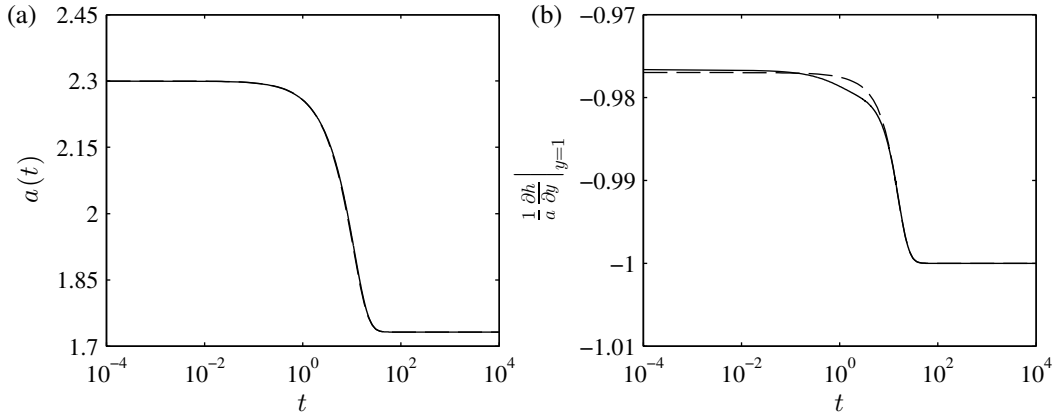}
    \caption{Interface formation model with $\chi_m=0$: \new{Evolution of (a) the droplet fronts and (b) the microscopic contact angle for initial droplet radius $a(0)=2.3$ to demonstrate a receding droplet situation with parameter values $\bar\beta=10^3$, $\bar\beta\bar\tau=1$, ${\bar\alpha}=1/12$, $\breve{\rho}=1$. Dashed lines use the asymptotic equations from matching with $c_0 = -0.1$ and $\hat{P}_0/\dot{a} = 0.54$, with solid lines from the solution of the PDEs.}} 
\label{fig:receding}
\end{figure}

Figures \ref{fig:origpde}--\ref{fig:receding} show the evolution of the droplet fronts and the \new{microscopic} dynamic contact angle behaviour in some instances for the $\chi_m=0$ interface formation model. Solid lines are from the full numerical solution of the partial differential equations, with dashed lines based upon the asymptotic results of the spreading rate in \refe{eq:adotina}, with $c_0$ determined through the solution of the ODEs in Sec.~\ref{sec:fullOadotchi0} \new{for the droplet radius, and with the microscopic contact angle $a^{-1}\pilfrac{h}{y}|_{y=1} = -(1+\hat{P}_0)$. Unlike for the Navier-slip case,\cite{Savva09} when considering spreading outside the the region of asymptotic validity (given in \refe{eq:c0adotvalid}), we are not always able to obtain agreement between asymptotic and full numerical solutions. As such, the initial droplet radius is always chosen such that \refe{eq:c0adotvalid} holds.}

Figure \ref{fig:origpde} has parameters chosen to be $\bar\beta=10^3$, $\bar\beta\bar\tau=1$, ${\bar\alpha}=1/12$, $\breve{\rho}=1$ as in Figs. \ref{fig:chi0odeprofile1}--\ref{fig:chi0odeprofile2} and the streamline plot in Fig.~\ref{fig:maran}(a) to show the excellent agreement between the asymptotic and full numerical results, \new{where these typical parameter values have $c_0=-0.1$ and $\hat{P}_0/\dot{a} = 0.54$ from our asymptotic results}. \new{Figure \ref{fig:bigc0chi0a0v16} has the extreme parameter values $\bar\beta=10^3$, $\bar\beta\bar\tau=100$, ${\bar\alpha}=1/12$, $\breve{\rho}=10$, which give $c_0 = 56.96$ and $\hat{P}_0/\dot{a} = 55.4$. Excellent agreement is seen for the droplet radius, and good agreement for the microscopic contact angle for larger times. We would not expect the microscopic contact angle comparisons to show perfect agreement, especially for these extreme parameter values, due to its dependence
on the contact line velocity and thus the initial condition prescribed. This can also be observed with slip
and precursor film models, see Fig.~1(c) of Ref.~\onlinecite{SavvaPrecursorSlip}.}

Figure \ref{fig:chi0asmaran} gives two further examples of the droplet radius evolution using the parameters of the streamline and velocity plots in Figs. \ref{fig:maran}(b) and \ref{fig:rollingvelplots}(a), being (a) $a(0)=1.3$, $\bar\beta=10^3$, $\bar\beta\bar\tau=0.1$, ${\bar\alpha}=1/12$, $\breve{\rho}=10$ and (b) $a(0)=1.2$, $\bar\beta=10^3$, $\bar\beta\bar\tau=1$, ${\bar\alpha}=1/12$, $\breve{\rho}=0.1$. Finally, Fig.~\ref{fig:receding} demonstrates a situation for a receding droplet. The initial droplet radius is \new{a(0)=2.3}, with parameter values chosen as $\bar\beta=10^3$, $\bar\beta\bar\tau=1$, ${\bar\alpha}=1/12$, $\breve{\rho}=1$, corresponding to $c_0 = -0.1$ \new{and $\hat{P}_0/\dot{a} = 0.54$}.

\subsection{Numerical solution of the full problem for $\chi_m=1$}
\label{sec:numchi1}

\begin{figure}[ht]
    \centering
    \includegraphics{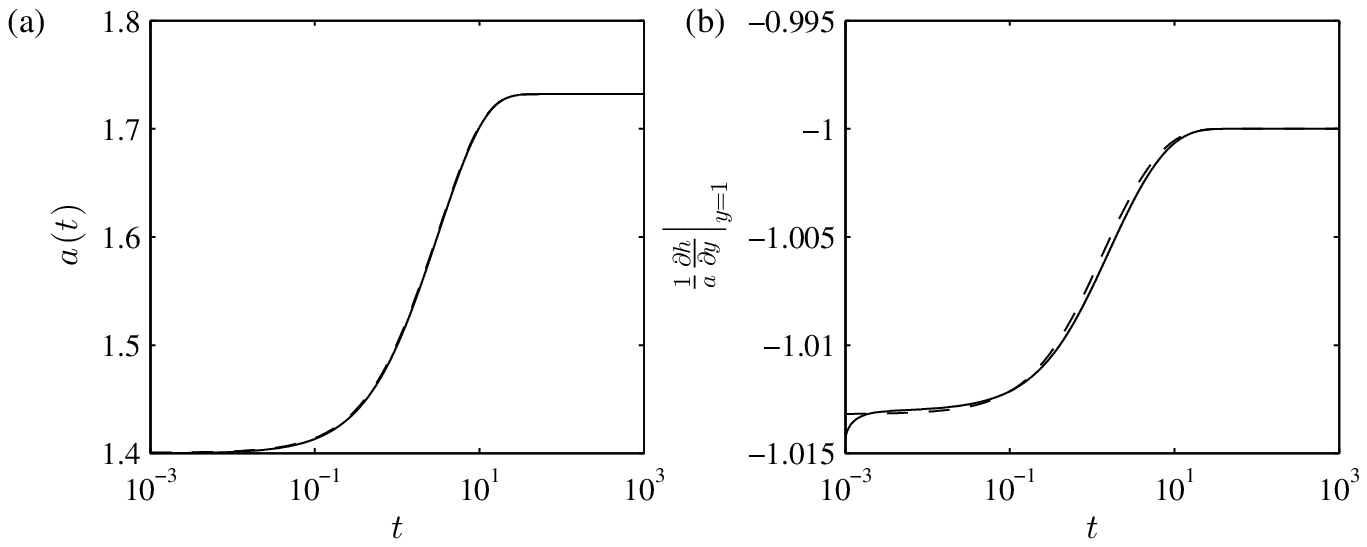}
    \caption{Interface formation model with $\chi_m=1$: \new{Evolution of (a) the droplet fronts, and (b) the microscopic contact angle for initial droplet radius $a(0)=1.4$ and parameter values $\bar\beta=10^3$, $\bar\beta\bar\tau=1$, ${\bar\alpha}=1/12$, $\breve{\rho}=0.1$, $\bar{R}_0=1/3$. Dashed lines use the asymptotic equations from matching with $c_0 = 0.12$ and $\hat{P}_0/\dot{a} = 0.093$, with solid lines from the full numerical solution.}}
\label{fig:chi1fplot}
\end{figure}

\begin{figure}[ht]
    \centering
    \includegraphics{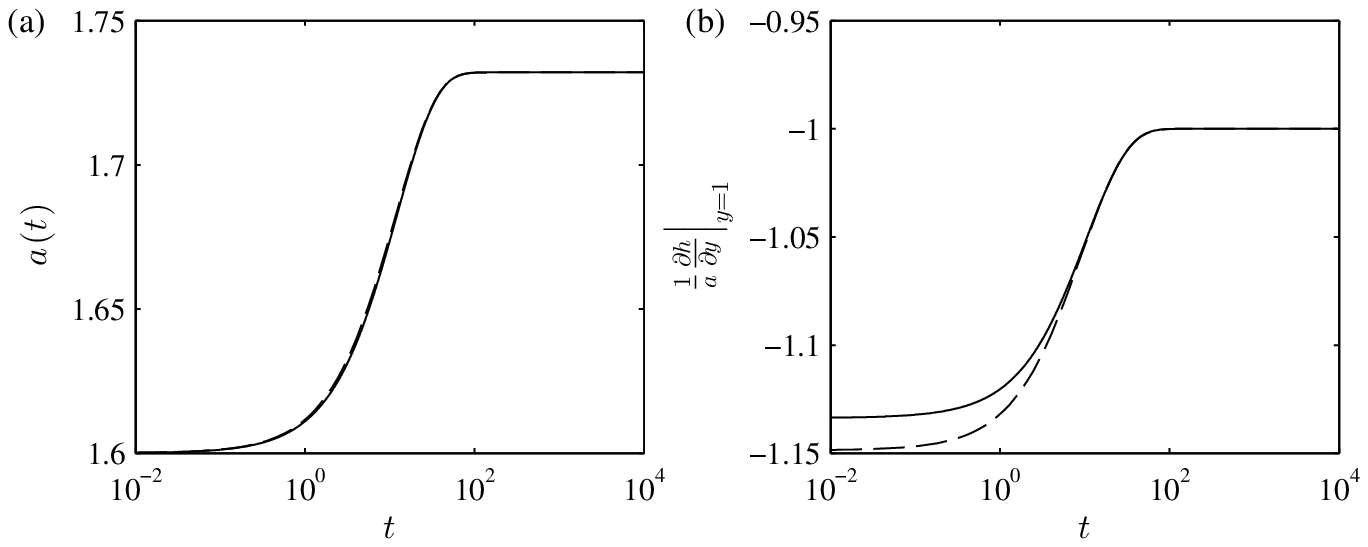}
    \caption{Interface formation model with $\chi_m=1$: \new{Evolution of (a) the droplet fronts, and (b) the microscopic contact angle for initial droplet radius $a(0)=1.6$ and parameter values $\bar\beta=10^3$, $\bar\beta\bar\tau=10$, ${\bar\alpha}=0.1$, $\breve{\rho}=10$, $\bar{R}_0=0.1$. Dashed lines use the asymptotic equations from matching with $c_0 = 10.44$ and $\hat{P}_0/\dot{a} = 12.09$, with solid lines from the full numerical solution.}}
\label{fig:biggerc0chi1}
\end{figure}

We now return to the governing equations of the full boundary value problem for $\chi_m=1$, \refe{eq:chi1eveq} and \refe{eq:chi0AB1}, \refe{eq:chi0AB2}, \refe{eq:chi0ocl3}--\refe{eq:chi0rhoSG}. Transforming the system to the non-moving coordinate $y$ through $\Bax = a y$, the governing equations become
\begin{equation}
 \pfrac{\Bah}{\Bat} - \frac{\dot{a}y}{a}\pfrac{\Bah}{y}
+ \frac{1}{a}\pfrac{}{y}\left(-\frac{\Bah^3}{2a^3} \pppfrac{\Bah}{y}
+ \frac{3}{2}A\Bah^2 + B\Bah\right) = {\bar{R}_0} \mathcal{D},
\label{eq:numchi1e1}
\end{equation}
with the remaining equations as for $\chi_m=0$, given in \refe{eq:eq2actualprob}--\refe{eq:outrgrs}, and with the modified contact line velocity $ \bar{U}_{CL} = B - a\bar R_0\left(\partial_y{\Bah}\right)^{-1}\mathcal{D}$. Considering \refe{eq:numchi1e1} as $y\to 1$, and implementing the second boundary condition, we require
\begin{equation}
\dot{a} = B(1) - \frac{\bar{R}_0}{2}\left( \frac{a}{\partial_y{\Bah}(1)}-\frac{\partial_y{\Bah}(1)}{a}\right),
\end{equation}
which confirms $\bar{U}_{CL}=\dot{a}$. The procedure follows that of the $\chi_m=0$ case closely, in particular the substitutions \refe{eq1:subsrem5th2} and \refe{eq:GasH}, and the conditions \refe{eq:numchi0eotherconds}, are employed, noting that \refe{eq:massconR0} and implementation of the contact line conditions gives the modified area condition:
\begin{equation}
a\int_0^1 h dy = 1 - \bar{R}_0\bar{\tau}(1+\breve\rho)(a-a(\infinity)).
\end{equation}
The mass of the droplet remains constant, but the area varies in the above condition as the proportion of the mass in the surface layers varies due to the mass transfer terms in the governing equations. Figures \ref{fig:chi1fplot} and \ref{fig:biggerc0chi1} show excellent agreement between asymptotic and full numerical solutions for the evolution of the droplet fronts, with the \new{microscopic} dynamic contact angle behaviours also shown. Initial droplet radii are chosen to ensure the conditions of our asymptotic analysis are satisfied. Figure \ref{fig:chi1fplot} depicts the evolution using the parameter values of the velocity plot in Fig.~\ref{fig:rollingvelplots}(b), being $\bar\beta=10^3$, ${\bar\alpha}=1/12$, $\breve{\rho}=0.1$, $\bar\beta\bar\tau=1$, $\bar{R}_0=1/3$, where $c_0 = 0.12$ \new{and $\hat{P}_0/\dot{a} = 0.093$}. Figure \ref{fig:biggerc0chi1} uses parameter values $\bar\beta=10^3$, $\bar\beta\bar\tau=10$, ${\bar\alpha}=0.1$, $\breve{\rho}=10$, $\bar{R}_0=0.1$ to compare results for the larger value, $c_0 = 10.44$ \new{and $\hat{P}_0/\dot{a} = 12.09$}.

%
%

\section{Conclusions}
\label{sec:conclusions} We have conducted a critical analysis of the theory of contact line motion proposed by Shikhmurzaev, in both original \cite{Shikh93} and modern \cite{ShikhBook} formulations through application to the problem of a thin two-dimensional droplet spreading on a flat solid substrate. We confirm a number of features of the model in this scenario, in particular:
\begin{itemize}
 \item The interface formation model is able to alleviate the logarithmic pressure singularity at the contact line, a problem encountered with the Navier-slip model. \new{(The Ruckenstein and Dunn slip model displays nonsingular pressure behaviour, as do other models mentioned in the Introduction, Sec.~\ref{sec:introduction}, such as those with diffuse interfaces or precursor films)}.
 \item For this particular flow scenario the model needs no adjustment. The requirement of finite pressure at the contact line closes the system and no additional contact line conditions are required.
 \item \new{If the logarithmic pressure singularity is allowed (given that it will still represent a finite force), then one may still adhere to the additional condition of Bedeaux\cite{BedeauxExtraCond} and Billingham,\cite{Bill06} but with a further model parameter to be imposed, alongside the myriad of parameters already necessary in the interface formation model.} 
 \item The \new{microscopic} contact angle is determined through the solution of the governing equations, with no assumptions required on its velocity dependence from experimental results.
\end{itemize}
It is instructive to make some further remarks about the model, \new{relevant to both situations of enforcing finite pressure at the contact line or instead applying the additional condition}:
\begin{itemize}
 \item For quasistatic droplet spreading \new{in the long-wave approximation} the interface formation model effectively reduces to a sophisticated slip model, with spreading rates based on a parameter $c_0$, obtained through our matching procedure.
\item An analytical formula for $c_0$ in terms of the model parameters has not been possible, requiring instead a numerical solution of a system of two ODEs.
\item There is a large number of parameters available to vary when compared to slip models. Even in the simple setting considered here, where assumptions have been made to keep them to a minimum (for instance that the gas-liquid and solid-liquid interfaces share the same values for $\tau$, $\gamma$, $\alpha$ and $\beta$). Uncertainty in their values, especially if the model is to be applied to a range of different fluids in a variety of technological process, makes predictions difficult. Detailed and systematic measurements of the parameters, either experimentally or through molecular dynamics simulations, are needed.
\item The complexity of the model, even when applied to the simple situation adopted here, \new{may be part of the reason} why there is not a larger body of research using it---\new{along with the questions about the physical basis of the model, discussed in Sec.~\ref{sec:introduction}}. As many of the crucial features of the model have remained in the long-wave approximation for the droplet spreading problem, the latter could serve as a useful starting point to consider further details of the model. On the other hand, the Ruckenstein and Dunn slip model seems to capture all the necessary physical ingredients of contact line dynamics with the exception of \new{nanoscale} rolling motion \new{(the experimental verification, or indeed prohibition, of which is beyond the capability of current results known to the authors)}. It could thus be used as the basis for the development of models which capture this effect also, \new{if desired}, and yet are simpler and with a smaller number of parameters to the interface formation model.
\end{itemize}

Taking full account of the mass transfer between bulk and surface layers, as in the modern formulation of the interface formation model, is required for \new{nanoscale} rolling motion through the contact line. The original formulation predicts a low-velocity region as seen for slip models, which prevents the fluid reaching the contact line in finite time. The evolution of the droplet radius using the interface formation model reduces to an equivalent expression for a slip model, with the \new{microscopic} dynamic contact angle having a correction to its static value proportional to the contact line velocity. This is also equivalent to the spreading rate for a slip model with \new{microscopic} dynamic contact angle equal to the static contact angle, but with a slip length modified by a parameter $c_0$, appearing in the outer expansion of the inner region equations.

Our study has been based on the simplest of settings, that of a droplet spreading on a planar horizontal substrate, where the dynamic behaviour of the interface formation model could be investigated, both analytically and numerically. The addition of substrate topography, chemical heterogeneities, gravitational effects and the extension to three dimensional droplets to allow for comparison with experiments would all be of interest (albeit increasing the complexity of the analysis), as such effects have been shown to induce many interesting and often surprising phenomena (see e.g. the recent studies in Refs.~\onlinecite{Savva09,RajChemHet,SavvaKalliadasisPavliotis,SavvaPavliotisKalliadasis1,SavvaPavliotisKalliadasis2}). The value of the relaxation time $\tau$ has been one of the points of contention for the interface formation model.\cite{EggersOnShikh} Provided an extension to three dimensional droplets produces similar results, an experimental comparison could be drawn for the evolution of the droplet radius. For known values of the other parameters, then $c_0$, the constant determining the rate of spread of the droplet, could be fitted to the experimental data which in turn could be used to determine $\tau$. It would then be of interest to compare this to the range of values suggested by Blake and Shikhmurzaev,\cite{BlakeDynWet} and motivated the typical values adopted here and by Billingham.\cite{Bill06,BillGrav}

%
%

\section*{Acknowledgements}
We are grateful to the anonymous referees for insightful comments and
suggestions. We acknowledge financial support from ERC Advanced
Grant No. 247031.

%
%

\appendix

\section{Simplified dimensional equations}

We record the simplified equations relevent to Sec.~\ref{sec:appdropge}, having eliminated excess variables. On the solid-liquid interface we have
\begin{gather}
 \pfrac{U}{Z} + \pfrac{V}{X} -\frac{\gamma}{2\mu}\pfrac{\rho_S^s}{X}=\frac{\beta}{\mu} U,\qquad
V = \chi_m\frac{\rho_S^s-\rho_{Se}^s}{\rho\tau},\nonumber\\
\pfrac{\rho_S^s}{T}+\pfrac{}{X}\left[
\frac{\rho_S^s{U}|_{Z=0}}{2}-\gamma\alpha\rho_S^s\pfrac{\rho_S^s}{X}\right]
= \frac{\rho_{Se}^s-\rho_S^s}{\tau},
\end{gather}
having eliminated ${u}_S^s$ and $\sigma_S^s$ using the last two equations of \refe{eq:2dsbc6}. On the gas-liquid interface we eliminate $\sigma_G^s$ using \refe{eq:2dlbc7}. It is then possible to eliminate both components of the surface velocity ${u}_{G}^s$ and ${v}_{G}^s$ using \refe{eq:2dlbc6} and \refe{eq:2dlbc4} to find
\begin{equation}
u_G^s
= U - \gamma\frac{1+4{\bar\alpha}}{4\beta\mathcal{H}} \pfrac{\rho_G^s}{X}-\chi_m\frac{\rho_G^s-\rho_{Ge}^s}{\tau\rho\sqrt{\mathcal{H}}}\pfrac{H}{X},
\qquad
v_G^s
=V- \gamma\frac{1+4{\bar\alpha}}{4\beta \mathcal{H}
}\pfrac{\rho_G^s}{X}\pfrac{H}{X}+\chi_m\frac{\rho_G^s-\rho_{Ge}^s}{\tau\rho\sqrt{\mathcal{H}}},
\end{equation}
then giving the surface equations
\begin{subequations}
\begin{align}
&\pfrac{H}{T}+U\pfrac{H}{X}=V
+\chi_m\frac{(\rho_G^s-\rho_{Ge}^s)\sqrt{\mathcal{H}}}{\tau\rho}
,\\
&P_G-P+\tbf{n}_G\bm{\cdot}\tbf{T}\bm{\cdot}\tbf{n}_G
=
\gamma\frac{\rho_{(0)G}^s-\rho_G^s}{\mathcal{H}^{3/2}}\ppfrac{H}{X}
,
\\
& \left( \mathcal{H}-2\right)\left( \pfrac{U}{Z}+\pfrac{V}{X} \right)+4\pfrac{U}{X}\pfrac{H}{X} =
\frac{\gamma\sqrt{\mathcal{H} }}{\mu} \pfrac{\rho_G^s}{X},
\\
&\pfrac{\rho_G^s}{T}
- \gamma(1+4{\bar\alpha})\frac{\mathcal{H}\pilfrac{}{X}\left(\rho_G^s\pilfrac{\rho_G^s}{X}\right) - \rho_G^s\pilfrac{\rho_G^s}{X}\pilfrac{H}{X}\ppilfrac{H}{X}
}{4\beta\mathcal{H}^2}
+\frac{\left[ V\pilfrac{H}{X}+U \right]\pilfrac{\rho_G^s}{X}}{ \mathcal{H} }
\nonumber\\
&- \chi_m\frac{ \rho_G^s- \rho_{Ge}^s}{ \tau\rho\mathcal{H}^{3/2} }\rho_G^s\ppfrac{H}{X}
+ \frac{\rho_G^s\left[ \pilfrac{H}{X}\pilfrac{}{X}(V|_{Z=H})+\pilfrac{}{X}(U|_{Z=H}) \right]}{ \mathcal{H} }
= \frac{\rho_{Ge}^s-\rho_G^s}{\tau},
\end{align}
\end{subequations}
where $\tbf{n}_G\bm{\cdot}\tbf{T}\bm{\cdot}\tbf{n}_G$ is as given in \refe{eq:normst}.
Finally at the contact line we have
\begin{subequations}
\begin{align}\nonumber
&  \rho_G^s\left[
   U - \frac{\gamma(1+4{\bar\alpha})}{4\beta\mathcal{H}}\pfrac{\rho_G^s}{X}
   -\chi_m\frac{\rho_G^s-\rho_{Ge}^s}{\tau\rho\sqrt{\mathcal{H}}}\pfrac{H}{X}
\vphantom{\pfrac{H}{X}}
   -{U}_{CL}\right]\cos\theta_d
   +\rho_G^s
   \left[ V - \frac{\gamma(1+4{\bar\alpha})}{4\beta\mathcal{H}}\pfrac{\rho_G^s}{X}\pfrac{H}{X}
\right.\nonumber\\&\left.
   +\chi_m\frac{\rho_G^s-\rho_{Ge}^s}{\tau\rho\sqrt{\mathcal{H}}} \right]\sin\theta_d
   + \rho_S^s\left[\frac{U}{2}-\gamma\alpha\pfrac{\rho_S^s}{X} -{U}_{CL}\right] = 0,
\\
&  (\rho_{(0)G}^s-\rho_G^s)\cos\theta_d-\rho_S^s= (\rho_{(0)G}^s-\rho_{Ge}^s)\cos\theta_s-\rho_{Se}^s,
\\
&   \rho_S^s\left[\frac{U}{2}-\gamma\alpha\pfrac{\rho_S^s}{X}
   -{U}_{CL}\right] = U_0 (\rho_S^s - \rho_{Se}^s + \rho_G^s - \rho_{Ge}^s),
\end{align}
\end{subequations}
having used $\sigma_{Ge}^s=\gamma(\rho_{(0)G}^s-\rho_{Ge}^s)$ and the equivalent in the solid surface. This completes our description of the governing equations for droplet motion, being able to eliminate the surface velocities and tensions, leaving a simplified set of equations.

%
%

\bibliography{clbib_July2012}
\bibliographystyle{unsrt}

\end{document}